\documentclass[useAMS,usenatbib]{mnras}
\usepackage{graphicx}
\usepackage{amsmath}
\usepackage{amssymb}
\usepackage{color}

\newcommand \etal {et~al.~}

\def \spose#1{\hbox  to 0pt{#1\hss}}  
\newcommand \lta{\mathrel{\spose{\lower 3pt\hbox{$\sim$}}\raise  2.0pt\hbox{$<$}}}
\newcommand \gta{\mathrel{\spose{\lower  3pt\hbox{$\sim$}}\raise 2.0pt\hbox{$>$}}}

\newcommand{\Lsun}{{\ifmmode{ {\rm L}_{\odot}}\else{L_{\odot}}\fi}}
\newcommand{\Msun}{{\ifmmode{ {\rm M}_{\odot}}\else{M_{\odot}}\fi}}
\newcommand{\Mstar}{{\ifmmode{M_{\rm star}}\else{$M_{\rm star}$}\fi}}
\newcommand{\Mhalo}{{\ifmmode{M_{\rm halo}}\else{$M_{\rm halo}$}\fi}}
\newcommand{\kms}{{\ifmmode{ {\rm km\,s^{-1}} }\else{ ${\rm km\,s^{-1}}$ }\fi}}
\newcommand{\fbar}{\ifmmode{f_{\rm bar}}\else{$f_{\rm bar}$}\fi}
\newcommand \Omegadm {\ifmmode \Omega_{\rm dm} \else $\Omega_{\rm dm}$ \fi}
\newcommand \Omegam {\ifmmode \Omega_{\rm m} \else $\Omega_{\rm m}$ \fi} 
\newcommand \Omegab {\ifmmode \Omega_{\rm b} \else $\Omega_{\rm b}$ \fi} 
\newcommand \OmegaL {\ifmmode \Omega_{\rm \Lambda} \else $\Omega_{\rm \Lambda}$\fi} 
\newcommand \Deltavir {\ifmmode \Delta_{\rm vir} \else $\Delta_{\rm vir}$ \fi}
\newcommand \rhocrit {\ifmmode \rho_{\rm crit} \else $\rho_{\rm crit}$ \fi}

\title[Convergence in the structure of CDM haloes]{NIHAO XXV: Convergence in the cusp-core transformation of cold dark matter haloes at high star formation thresholds}
\author[Dutton \etal]{Aaron A. Dutton$^1$\thanks{dutton@nyu.edu},
  Tobias Buck$^2$, Andrea V. Macci\`{o}$^{1,3,4}$, 
  Keri L. Dixon$^{1,4}$, 
  \newauthor{Marvin Blank$^{1,4,5}$, Aura Obreja$^{6}$}\\
$^1$New York University Abu Dhabi, PO Box 129188, Saadiyat Island, Abu Dhabi, United Arab Emirates\\
$^2$Leibniz-Institut f\"{u}r Astrophysik Potsdam (AIP), An der Sternwarte 16, D-14482 Potsdam, Germany\\
$^3$Max Planck Institut f\"{u}r Astronomie, K\"{o}nigstuhl 17, 69117 Heidelberg, Germany\\
$^4$Center for Astro, Particle and Planetary Physics (CAP$^3$), New York University Abu Dhabi, United Arab Emirates\\
$^5$Institut f\"{u}r Theoretische Physik und Astrophysik, Christian-Albrechts-Universit\"{a}t zu Kiel, Leibnizstr. 15, D-24118 Kiel, Germany\\
$^6$University Observatory Munich, Scheinerstra\ss e 1, D-81679 Munich, Germany\\
}

\begin{document}

\date{Accepted 2020 September 28. Received 2020 September 28; in original form 2020 May 10}
\maketitle

\label{firstpage}

\begin{abstract}
We use cosmological hydrodynamical galaxy formation simulations from
the NIHAO project to investigate the response of cold dark matter
(CDM) haloes to baryonic processes. Previous work has shown that the
halo response is primarily a function of the ratio between galaxy
stellar mass and total virial mass, and the density threshold above
which gas is eligible to form stars, $n [{\rm cm}^{-3}]$.  At low $n$
all simulations in the literature agree that dwarf galaxy haloes are
cuspy, but at high $n\gta 100$ there is no consensus.  We trace halo
contraction in dwarf galaxies with $n\gta 100$ reported in some
previous simulations to insufficient spatial resolution.  Provided the
adopted star formation threshold is appropriate for the resolution of
the simulation, we show that the halo response is remarkably stable
for $n\gta 5$, up to the highest star formation threshold that we
test, $n=500$.  This free parameter can be calibrated using the
observed clustering of young stars. Simulations with low thresholds
$n\le 1$ predict clustering that is too weak, while simulations with
high star formation thresholds $n \gta 5$, are consistent with the
observed clustering.  Finally, we test the CDM predictions against the
circular velocities of nearby dwarf galaxies. Low thresholds predict
velocities that are too high, while simulations with $n\sim 10$
provide a good match to the observations.  We thus conclude that the
CDM model provides a good description of the structure of galaxies on
kpc scales provided the effects of baryons are properly captured.
\end{abstract}

\begin{keywords}
cosmology: theory -- dark matter -- galaxies: formation -- galaxies:
kinematics and dynamics -- galaxies: structure -- methods: numerical
\end{keywords}

\section{Introduction}

The structure of dark matter haloes on kiloparsec-scales provides a
sensitive astrophysical test of the standard cold dark matter
(CDM) paradigm, and more generally the nature of dark matter
\citep[e.g.,][]{Bullock17}.  Through the use of dissipationless
simulations, the structure of CDM haloes in the absence of baryons is
well determined \citep[e.g.,][]{Stadel09,Dutton14}.  The dissipation of gas to
the center of haloes is thought to only make the dark matter halo contract
\citep{Blumenthal86,Gnedin04}. However, other baryonic processes can
cause the dark matter halo to expand: dynamical friction from
infalling baryonic clumps \citep{El-Zant01}, resonances with galactic
bars \citep{Weinberg02}, and multiple episodes of gas outflows
\citep{Read05,Pontzen12,Dutton16b}.

Using 10 cosmological galaxy formation simulations from the MAGICC
project \citep{Stinson13}, \citet{DiCintio14} found that the structure
of CDM haloes, and hence the trade off between gas inflows and
outflows, depends on the ratio between the galaxy stellar mass and the
halo mass $\Mstar/\Mhalo$ (which is proportional to the integrated
star formation efficiency). At low $\Mstar/\Mhalo \lta 0.0003$ the
dark matter profile remains unchanged, due to minimal gas dissipation,
and minimal gas outflows.  As the efficiency increases the halo
expansion gets stronger, while the contractive effect of inflows still
remains small. The maximum expansion is reached at $\Mstar/\Mhalo\sim
0.003$. At higher $\Mstar/\Mhalo$ the expansion is reduced due to the
increasing importance of gas inflows until above $\Mstar/\Mhalo\sim
0.03$ the halo contracts.

This result was confirmed by \citet{Tollet16} using 60 simulations
from the NIHAO project \citep{Wang15}, and \citet{Chan15,Bullock17,Lazar20}
using simulated galaxies from the FIRE \citep{Hopkins14}
and FIRE-2 projects \citep{Hopkins18}. See also \citet{Maccio20} for
an extension to massive galaxies with AGN feedback.  However,
\citet{Bose19} finds essentially no change in the dark matter halo for
a wide range of $\Mstar/\Mhalo$ using the simulations from the AURIGA
\citep{Grand17} and APOSTLE \citep{Sawala16} projects.  The solution
to this apparent contradiction is that the halo response is strongly
dependent on the star formation density threshold adopted in the
simulation \citep{Governato10,Dutton19c, Benitez19}.  High thresholds $n\gta 10$
can result in halo expansion (for suitable $\Mstar/\Mhalo$), while
low thresholds $n\sim 0.1$ (e.g., APOSTLE, AURIGA) never result in
significant halo expansion.  While different simulation codes agree on
the halo response at low $n\sim 0.1$, there is not yet consensus for
high $n\gta 100$ with \citet{Benitez19} finding dark halo contraction
in dwarf galaxies, while FIRE, FIRE-2, \citet{Governato10, Governato12} and
\citet{Read16} finding halo expansion.

In order for hydrodynamical simulations of galaxy formation to make a
robust prediction for the structure of CDM haloes we thus need to find
ways to distinguish between simulations with different star formation
thresholds.  \citet{Dutton19c} showed that simulations with different
star formation thresholds resulted in differences in the variability
in the star formation rates, with a factor $\sim 2$ more variability
for $n=10$ than $n=0.1$.  \citet{Buck19b} showed that simulations with
different star formation thresholds resulted in differences in the
spatial distribution of young stars, specifically the two-point
correlation. Comparing to observations of the two-point correlation of
young star clusters from LEGUS \citep{Grasha17}, \citet{Buck19b} finds
$n=10$ consistent with observations, and $n=1$ and $n=0.1$
inconsistent, with clustering that is too weak.

In this paper we extend the studies of \citet{Dutton19c} and
\citet{Buck19b} to an order of magnitude higher star formation
thresholds with the goal of resolving the conflicting results at
$n\gta 100$. We also include a new set of simulations for $n=0.1$ that
we have recalibrated using the star formation efficiency, instead of
the feedback efficiency.

This paper is organized as follows: The simulation suite is outlined
in Section 2. Results on the convergence of the inner structure of CDM
haloes are given in Section 3. In Section 4 we constrain the star
formation threshold using the clustering of young stars. In Section 5
we test the CDM predictions with galaxy kinematics, finally a summary
is given in Section 6.

\begin{figure*}
  \includegraphics[width=0.8\textwidth]{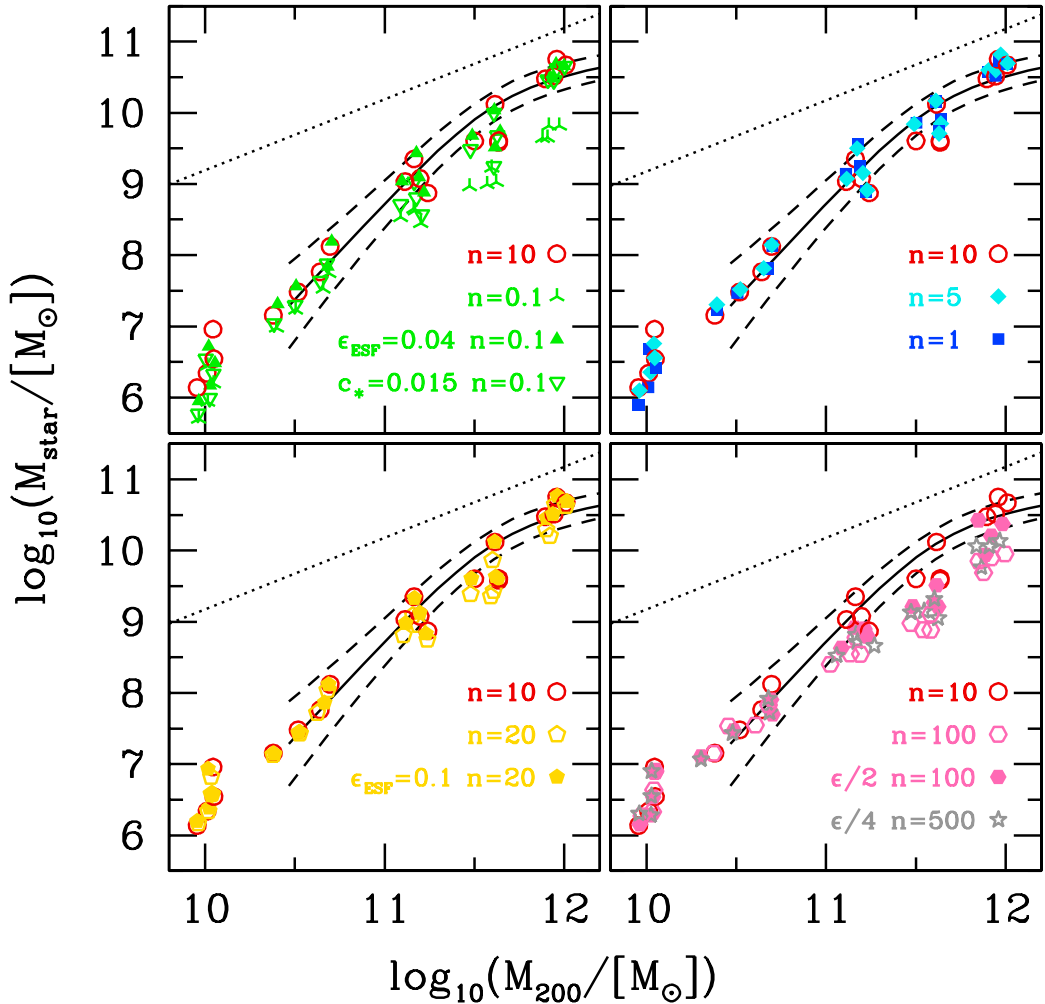}
  \caption{Relation between galaxy stellar mass, $\Mstar$, and total
    virial mass, $M_{200}$, at redshift $z=0$. The solid and dashed
    lines show results (mean and 1$\sigma$ scatter) from halo
    abundance matching \citep{Moster18}, while the dotted line
    corresponds to the cosmic baryon fraction. Simulations with
    different star formation thresholds, $n$, are given in different
    colors: $n=10$ (red, NIHAO default), $n=0.1$ (green), $n=1$
    (blue), $n=5$ (cyan), $n=20$ (yellow), $n=100$ (magenta), and
    $n=500$ (grey). Some simulations have been recalibrated by varying
    the star formation efficiency, $c_{\ast}$, or the early stellar
    feedback efficiency, $\epsilon_{\rm ESF}$, with new parameters as
    indicated.  At the highest $n$ we include simulations with smaller
    force softenings: $\epsilon/2$, and $\epsilon/4$.}
\label{fig:msmv}
\end{figure*}

\section{Simulations}
\label{sec:sims}

As in \citet{Dutton19c}, we use a set of 20 haloes of virial masses
between $M_{200} \sim 10^{10}$ and $\sim 10^{12}~\Msun$ taken from the
NIHAO project \citep{Wang15}. NIHAO is a sample of $\simeq 100$
hydrodynamical cosmological simulations run using the SPH code {\sc
  gasoline2} \citep{Wadsley17}.  The uniqueness of NIHAO is in the
combination of high spatial resolution over a wide range of halo
masses ($10^{10}$ to $10^{12}~\Msun$) for a statistical sample of
haloes.

The masses and force softenings of the dark matter particles are
chosen to resolve the mass profile at $\lta 1$ per cent of the virial
radius, which results in $\sim 10^6$ dark matter particles inside the
virial radius of all main haloes at $z=0$. The corresponding masses
and force softenings for the gas particles  are a factor of
$\Omegab/\Omegadm=0.182$ and $\sqrt{\Omegab/\Omegadm}=0.427$ lower.
Each hydro simulation has a corresponding simulation of the same
resolution, but with just dark matter particles (dark matter only,
DMO) of the same resolution. These DMO simulations have been started using
the identical initial conditions, replacing baryonic particles with
dark matter particles.

As discussed in detail in previous papers, and as outlined below,
NIHAO galaxies are consistent with a wide range of galaxy properties.
They form the right amount of stars (as compared to halo abundance
matching) both today and at earlier times \citep{Wang15}.  The masses
and half-light sizes of the cold gas are consistent with observations
\citep{Stinson15, Maccio16}. They follow several fundamental kinematic
scaling relations: the gas, stellar, and baryonic Tully-Fisher
relations \citep{Dutton17}, and the radial acceleration relation
\citep{Dutton19b}.  They match the clumpy morphology seen in observed
galaxies at high redshifts \citep{Buck17}.  They reconcile the
conflict between the steep halo velocity function of LCDM and the
shallow H{\sc i} linewidth velocity function observed in the nearby
Universe \citep{Maccio16, Dutton19a}.  They result in satellite mass
functions resembling the one of the Milky Way \citep{Buck19a}, and
emulate the Milky Ways central stellar bar \citep{Buck18, Buck19c}.
Given all of this success, they provide a good template with which to
predict the structure of cold dark matter haloes.

We refer the reader to \citet{Wang15} for a description of the NIHAO
simulations and \citet{Dutton19c} for more properties of the 20
resimulated galaxies. Here we briefly describe the parameters that we
vary in this paper: the star formation threshold $n$, the efficiency
of early stellar feedback, $e_{\rm ESF}$, and the star formation
efficiency, $c_{\ast}$.

In our simulations star formation is implemented as described in
\citet{Stinson06, Stinson13}.  Stars form from gas that is both cool
($T < 15 000$K) and dense ($\rho > n$[cm$^{-3}$]).  Gas that passes
both thresholds is converted into stars according to 
\begin{equation}
  \frac{\Delta\Mstar}{\Delta t} = c_{\ast} \frac{M_{\rm gas}}{t_{\rm dyn}}.
\end{equation}
Here $\Delta\Mstar$ is the mass of stars formed, $\Delta t=0.84\,$Myr
is the time-step between star formation events (Age of Universe$/2^{14}$), and $t_{\rm dyn}$ is
the gas particle's dynamical time. The fiducial efficiency of star
formation is set to $c_{\ast}=0.1$.

In our fiducial NIHAO simulations we adopt a star formation threshold
of $n=10\, [{\rm cm}^{-3}]$. This is chosen as it is roughly the
maximum density that we can resolve:
\begin{equation}
\label{eq:nmax}  
  n_{\rm max} \approx 50 \,m_{\rm gas}/\epsilon_{\rm gas}^3, 
\end{equation}
  where $m_{\rm gas}$ is the gas particle mass and $\epsilon_{\rm gas}$ is the gas
gravitational force softening. Here 50 is the number of SPH particles
in the smoothing kernel.  For all our simulations (with different
resolution levels) this formula results in the same $n_{\rm max}\approx 10$.

In this paper we present new results for simulations with $n=5$ and
$n=20$, as well as much higher values $n=100$ and $n=500$.  As we will
show below, in order to form galaxies with realistic baryonic mass
distributions with $n\gg 10$ we need smaller force softenings. Smaller
softening allows the gas to clump on smaller scales and thus reach
higher densities (Eq.~\ref{eq:nmax}).

The other parameter of relevance to this study is the feedback
efficiency.  The NIHAO simulations employ thermal feedback in two
epochs as described in \citet{Stinson13}. The first epoch models the
energy input from stellar winds and photoionization from bright young
stars before supernovae explode. We thus term this early stellar
feedback (ESF).  The ESF consists of a fraction $\epsilon_{\rm ESF}$
of the total stellar flux being ejected from stars into surrounding
gas ($2 \times 10^{50}$ erg of thermal energy per $\Msun$ of the
entire stellar population). Radiative cooling is left on for the ESF.

The second epoch models the energy input from supernovae and starts 4
Myr after the star forms.   Stars with mass $8 ~\Msun < M_{\ast} < 40
~\Msun$ eject both energy ($\epsilon_{\rm SN}\times 10^{51}$ erg/SN)
and metals into the interstellar medium gas surrounding the region
where they formed. Supernova feedback is implemented using the
blastwave formalism described in \citet{Stinson06}. To correct for
numerical radiative losses, this model applies a delayed cooling
formalism for particles inside the blast region for $\sim 30$ Myr. 

The default parameters of the feedback model are
$\epsilon_{\rm ESF}=0.13$ and $\epsilon_{\rm SN}=1.0$. They were calibrated
against the evolution of the stellar mass versus halo mass relation
from halo abundance matching \citep{Behroozi13,Moster13} for a $z=0$
Milky Way mass halo $\sim 10^{12}~\Msun$. In this paper we leave
$\epsilon_{\rm SN}=1.0$ and vary $\epsilon_{\rm ESF}$.

\subsection{Haloes and galaxies}

Haloes are identified using the MPI+OpenMP hybrid halo finder
\texttt{AHF}\footnote{http://popia.ft.uam.es/AMIGA} \citep{Gill04,
  Knollmann09}. The virial mass of each halo is defined as the mass of
all particles within a sphere whose average density is 200 times the
cosmic critical matter density, $\rhocrit=3H_0^2/8\pi G$.  For the
hydro simulations the virial mass, size and circular velocity of the
haloes are denoted: $M_{200}$, $R_{200}$, $V_{200}$.  For the DMO
simulations the corresponding properties are denoted with a
superscript, ${\rm DMO}$.  We define the stellar mass, $M_{\rm star}$,
of a galaxy to be the mass of stars   enclosed within spheres of
radius $r_{\rm gal}=0.2R_{200}$,  corresponding to $\sim 10$ to $\sim
50$ kpc.  consider the main halo per zoom-in simulation.

\begin{figure*}
  \includegraphics[width=0.8\textwidth]{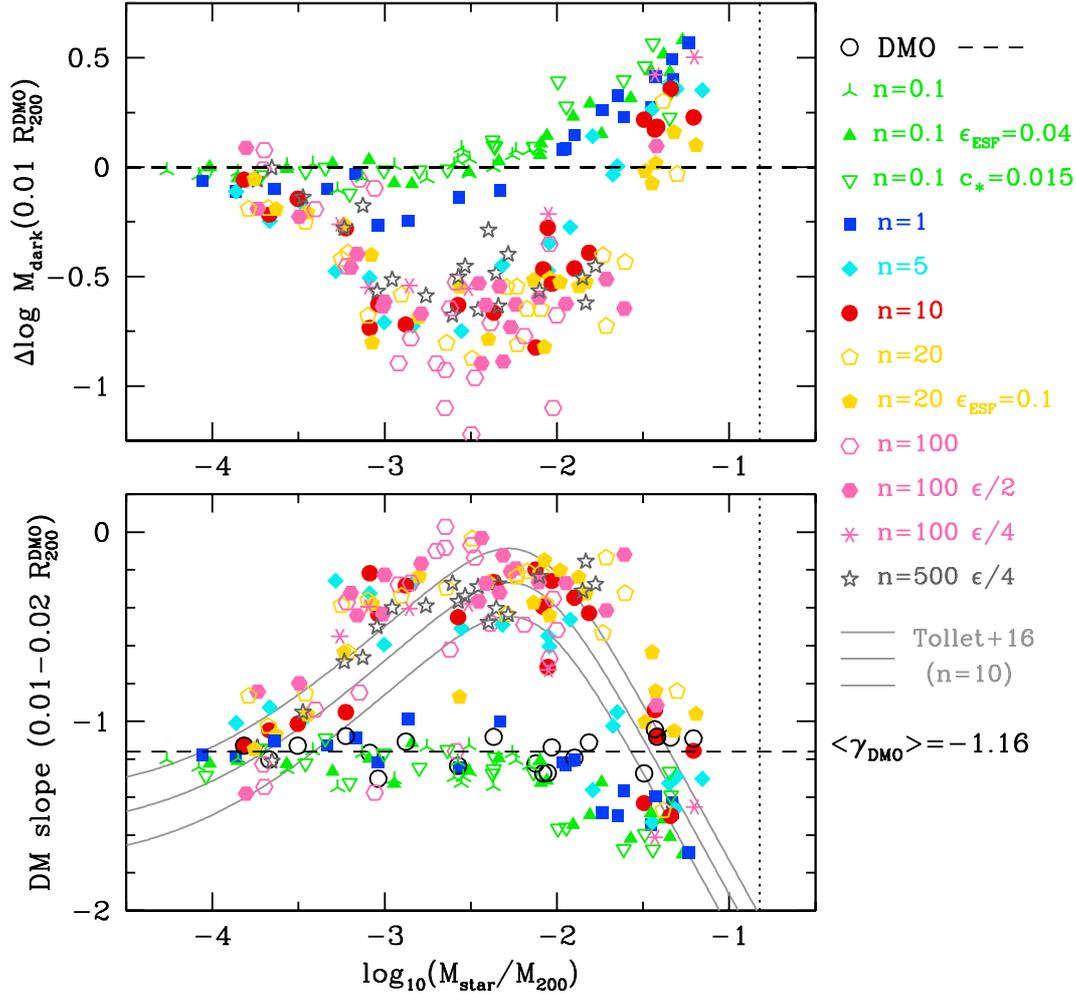}
 \caption{Change in enclosed mass ($\Delta\log M_{\rm dark} \equiv
   \log M_{\rm dark}^{\rm hydro} -\log M_{\rm dark}^{\rm DMO}$) at 1
   per cent of the virial radius (upper panels) and slope of the
   enclosed dark matter density profile between 1 and 2 per cent of
   the virial radius (lower panels) versus stellar to halo mass
   ratio. The dotted vertical line corresponds to the cosmic baryon
   fraction. The grey lines in the lower panel show the result from
   the full NIHAO sample from \citet{Tollet16}.  The star formation
   threshold varies from $n=0.1$ to $n=500$ with colors and symbols as
   indicated.  The default parameters are: efficiency of early stellar
   feedback $\epsilon_{\rm ESF}=0.13$, star formation efficiency
   $c_{\ast}=0.1$.  The halo response converges for $n\gta 5$ provided
   the force softening is small enough.}
\label{fig:alpha}
\end{figure*}

\subsection{Feedback recalibration}
Fig.~\ref{fig:msmv} shows the ratio between stellar and virial mass at
redshift $z=0$ for the 20 main haloes.  The solid (and dashed) lines
are the mean (and scatter) from halo abundance matching from
\citet{Moster18}, which we have corrected to our halo mass definition.
Each panel shows results for $n=10$ (fiducial NIHAO, red open circles)
together with one or two other values of $n$. For $n=1, 5$ and $10$
the relation is very similar (top right panel). However, for both
higher and lower $n$ the fiducial simulation parameters under produce
stars, especially so for the higher mass galaxies ($M_{200} >
10^{11}~\Msun$).  

For $n=0.1$ the fiducial parameters ($\epsilon_{\rm ESF}=0.13,
c_{\ast}=0.1$) result in stellar masses being under produced by an
order of magnitude in the $M_{200}\sim 10^{12}~\Msun$ haloes.  We
re-calibrate the $n=0.1$ simulations in two ways: by reducing the
efficiency of early-stellar feedback to $\epsilon_{\rm ESF}=0.04$
(filled triangles), or by reducing the star formation efficiency from
$c_{\ast}=0.1$ to $c_{\ast}=0.015$ (open inverted triangles).  The
latter effect may seem counter-intuitive, since without feedback a
lower $c_{\ast}$ is expected to result in a lower stellar
mass. However when feedback is included the opposite trend occurs.
Feedback has the effect of heating gas and removing it from star
forming regions, and thus naturally delays and reduces the amount of
stars formed.  Since denser gas radiates energy away faster, feedback
is less efficient when the surrounding gas is denser.  A lower
$c_{\ast}$ initially results in less star formation but also less
feedback energy injected into the ISM, which results in denser gas.
The denser gas reduces the efficiency of subsequent feedback events,
and thus results in more gas available to form stars, and ultimately
higher stellar masses.

For $n=20$ the reduction in stellar mass is relatively small. We
recalibrate by reducing the efficiency of early stellar feedback from
$\epsilon_{\rm ESF}=0.13$ (open yellow pentagons) to $\epsilon_{\rm
  ESF}=0.10$ (filled yellow pentagons).  The $n=100$ simulations have
a similar under production of stars as the $n=0.1$ simulations. We
have experimented with varying the feedback efficiency with limited
success. The problem for these simulations is more fundamental than
the choice of model parameters: they simply lack the spatial
resolution to enable sufficient amounts of gas to locally reach the
star formation threshold.  For $n=100$ we run simulations where all
particles have half the force softening $\epsilon/2$ (filled magenta
hexagons). For $n=500$ we use one quarter of the standard softening
$\epsilon/4$ (grey stars).  The main effect of reducing the force
softening is that it allows the smoothed gas densities to be higher
(since we set the SPH smoothing length to be proportional to the
gravitational force softening).  A factor of 2 lower force softening
thus allows a factor of 8 denser gas, and a factor of 4 lower force
softening allows a factor of 64 denser gas.  These haloes form
significantly more stars than the standard softening simulations, yet
they still under produce stars (in haloes with $M_{200}>
10^{11}~\Msun$) relative to the $n=10$ simulations and halo abundance
matching. So these simulations would benefit with further
re-calibration of the feedback efficiency and/or star formation
efficiency.  We do note though that for dwarf galaxy haloes
$10^{10}\lta M_{200} \lta 10^{11}~\Msun$ the stellar masses are
remarkably insensitive to the star formation threshold and choice of
force softening.

\begin{figure*}
  \includegraphics[width=0.45\textwidth]{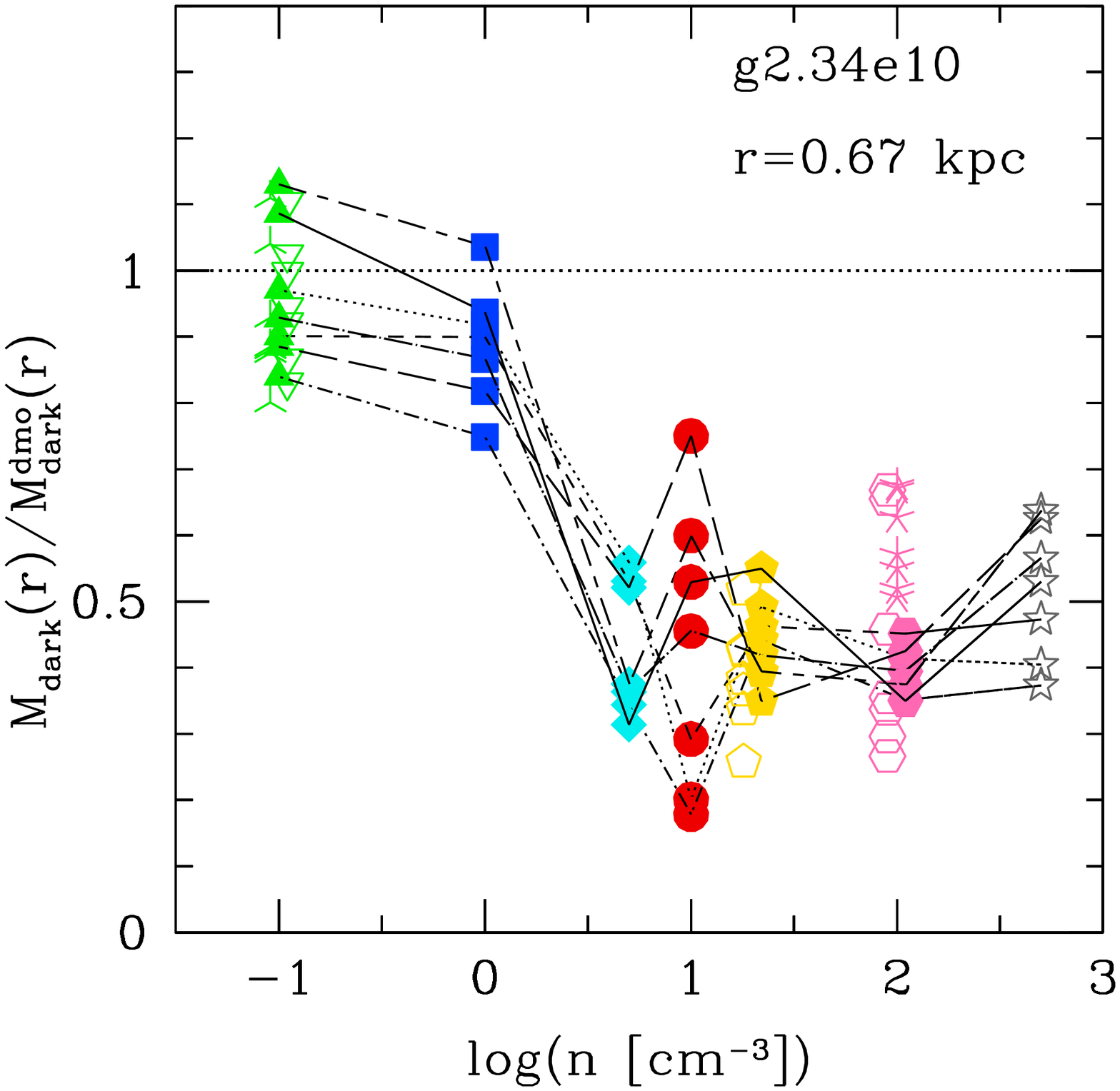}    
  \includegraphics[width=0.45\textwidth]{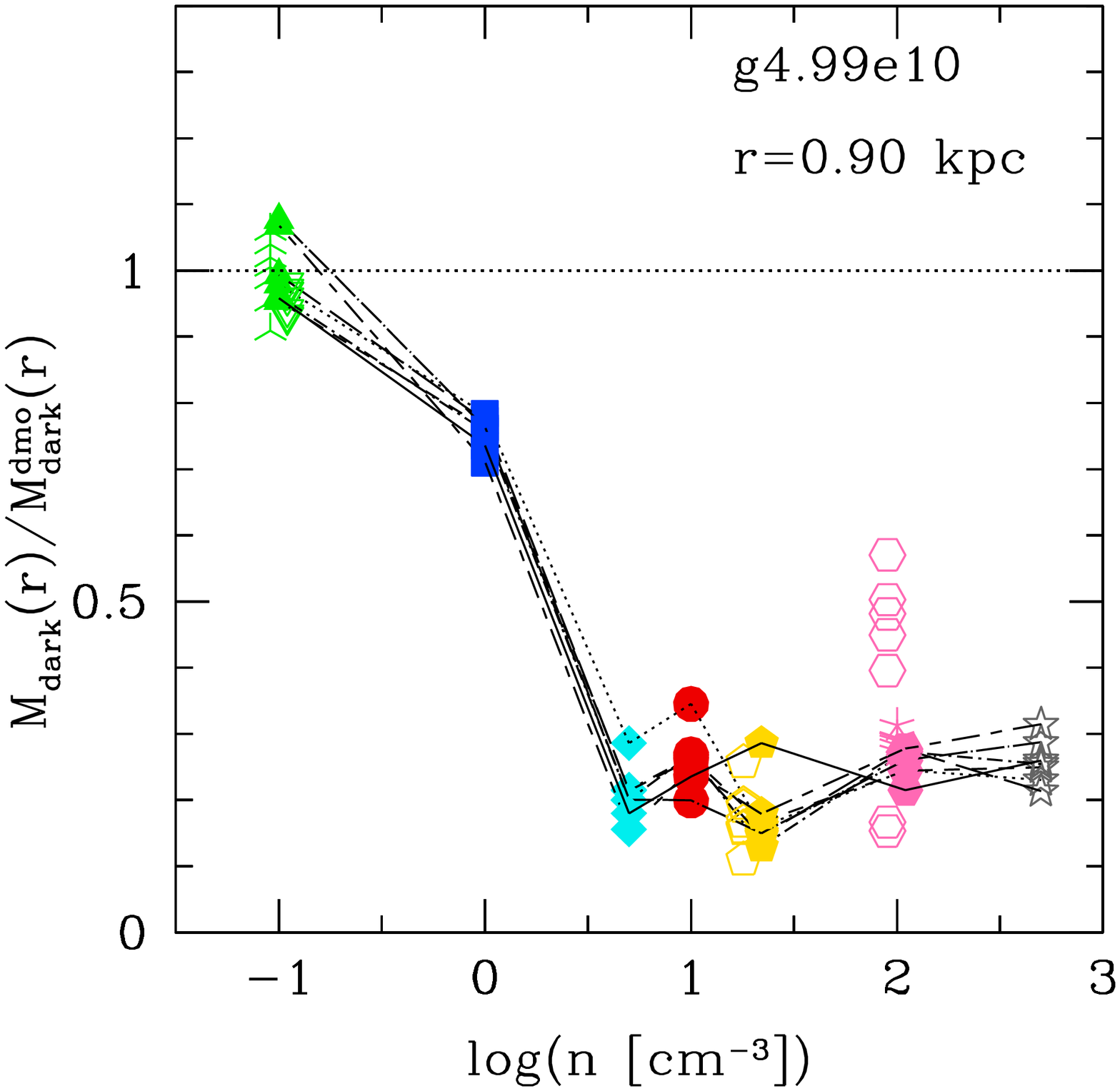}    
  \includegraphics[width=0.45\textwidth]{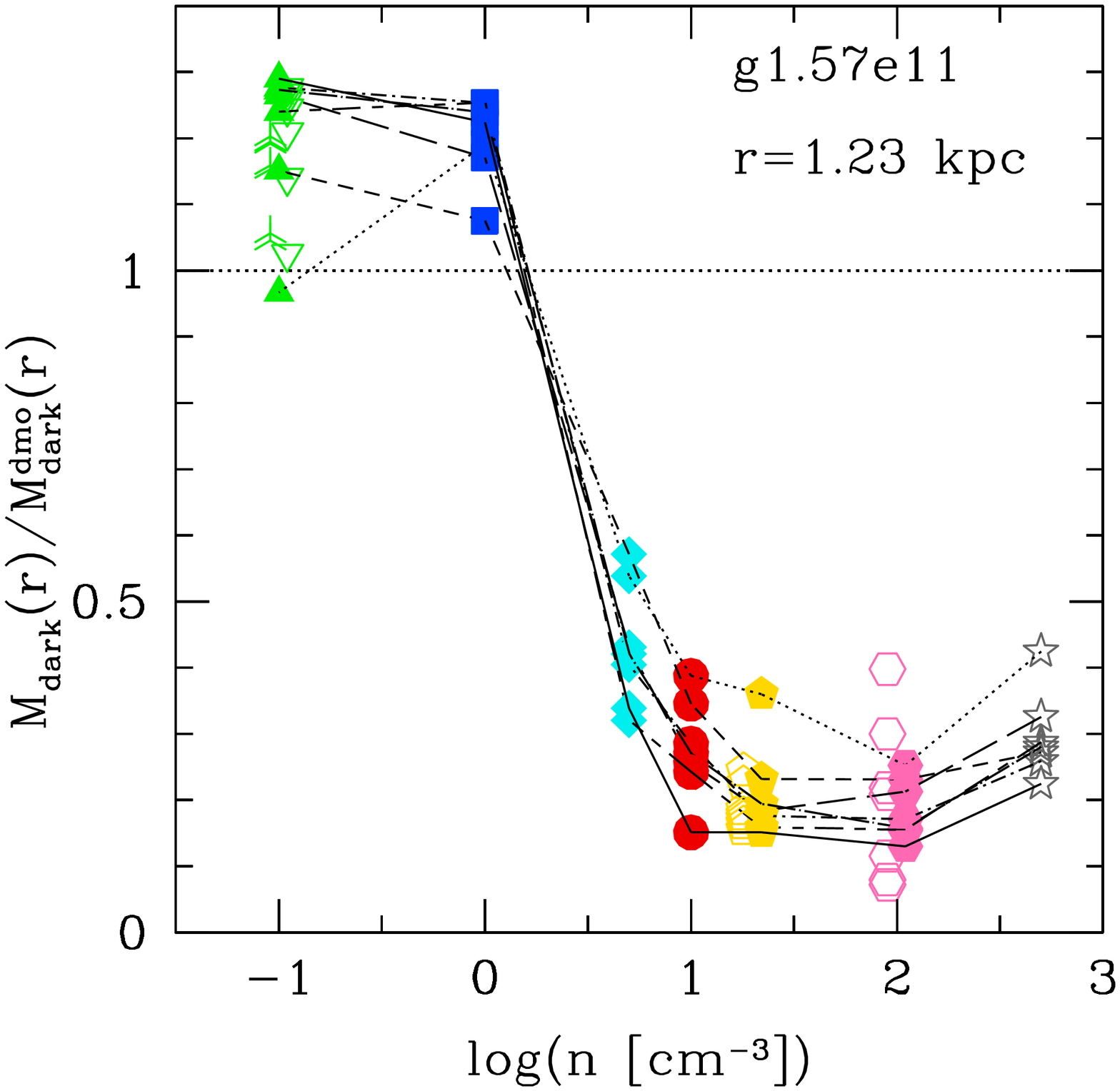}
  \includegraphics[width=0.45\textwidth]{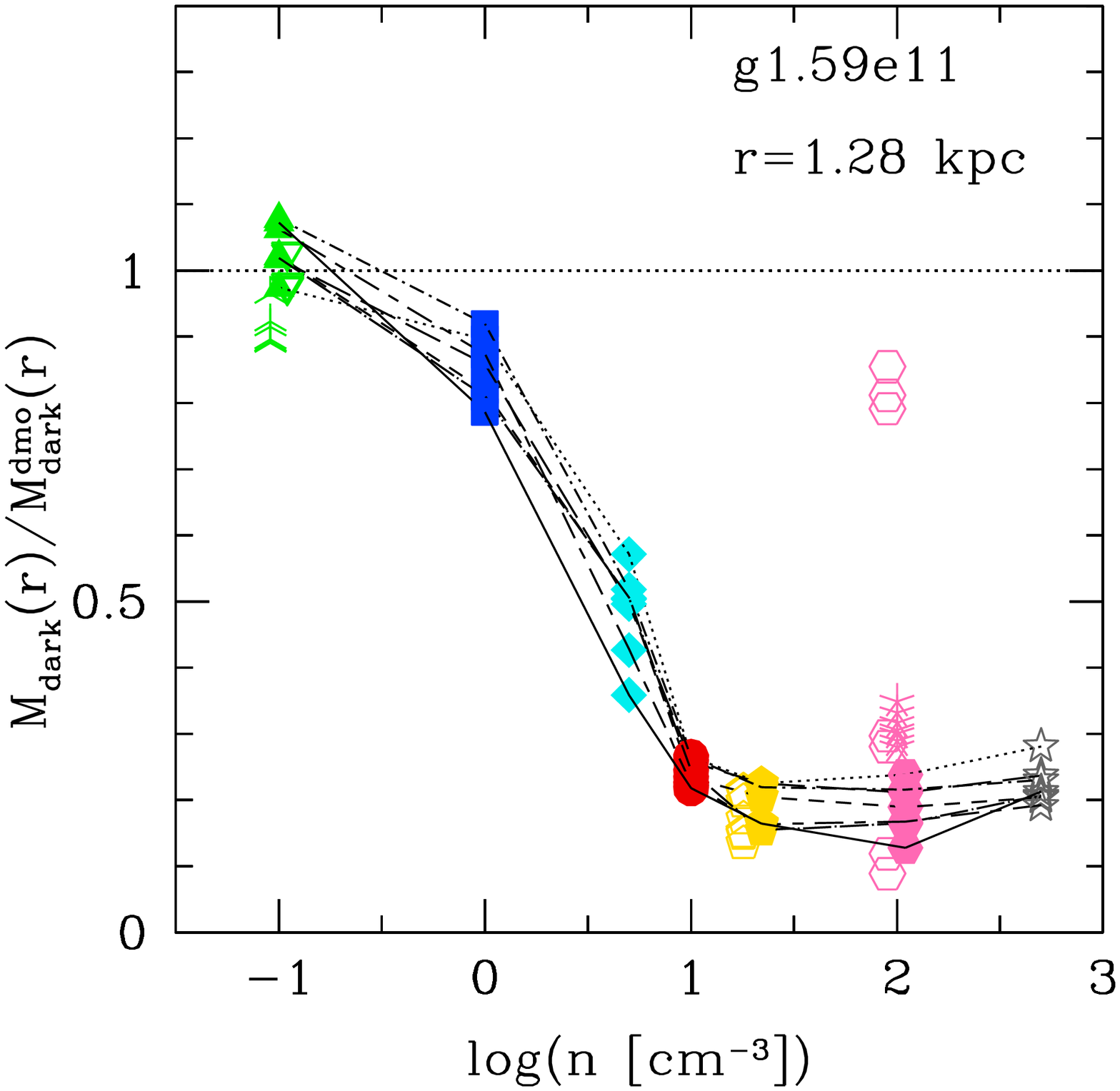}
  \caption{Ratio between dark matter mass in hydro and DMO simulations
    measured at 1 per cent of the virial radius versus the star
    formation threshold of the simulation. Point types and colors are
    as in Fig.~\ref{fig:alpha}. For thresholds with more than one set
    of simulations ($n=0.1, n=20, n=100$) we have introduced small
    horizontal offsets for clarity.  For each simulation we show 7
    outputs equally spaced in time between redshifts $z=0.5$ and
    $z=0.0$. The lines connect simulations at a given redshift, where
    solid is $z=0$. The average halo response converges for $n\gta
    10$.}
\label{fig:deltam}
\end{figure*}

\begin{figure*}
  \includegraphics[width=0.45\textwidth]{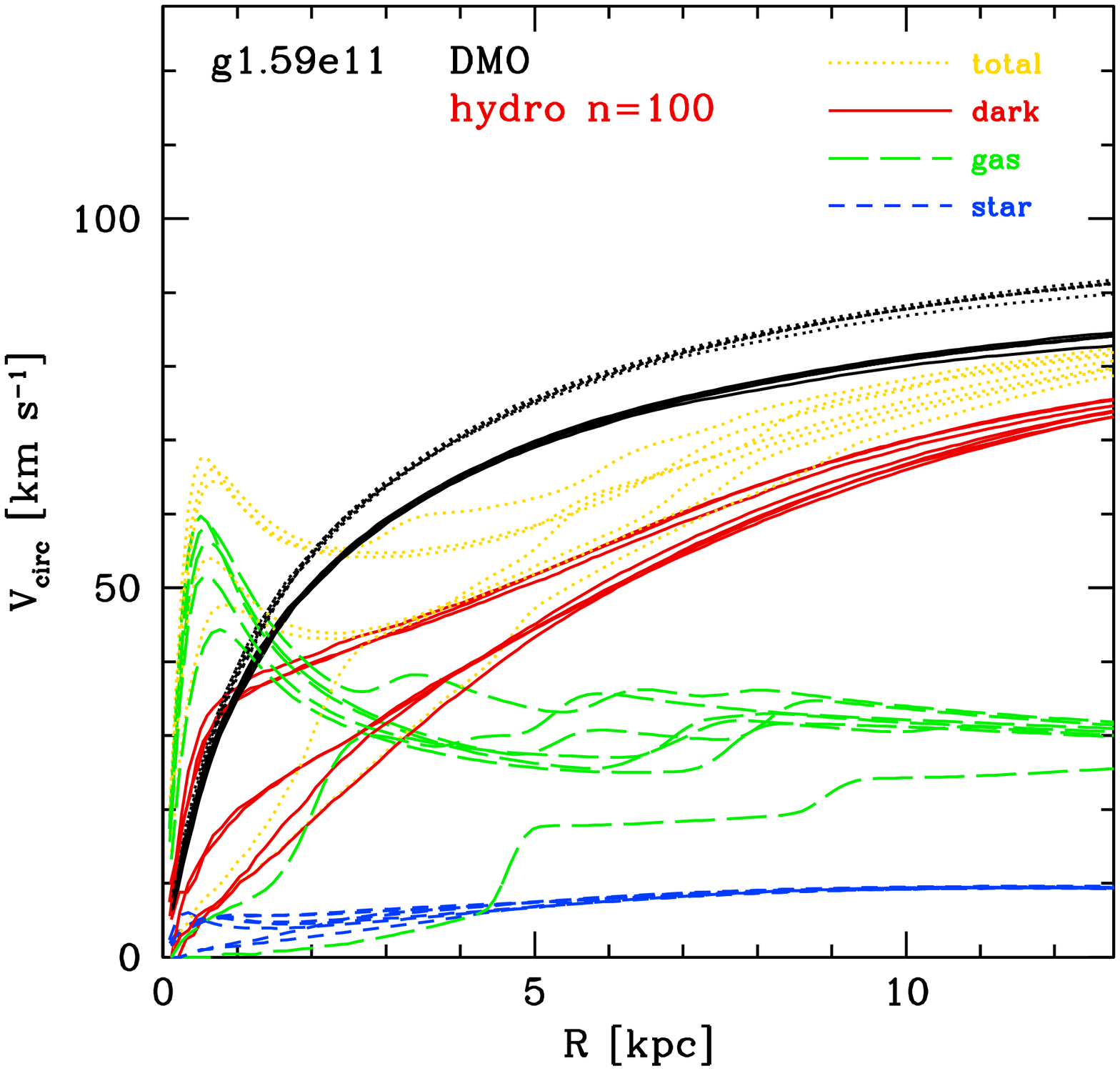}
  \includegraphics[width=0.45\textwidth]{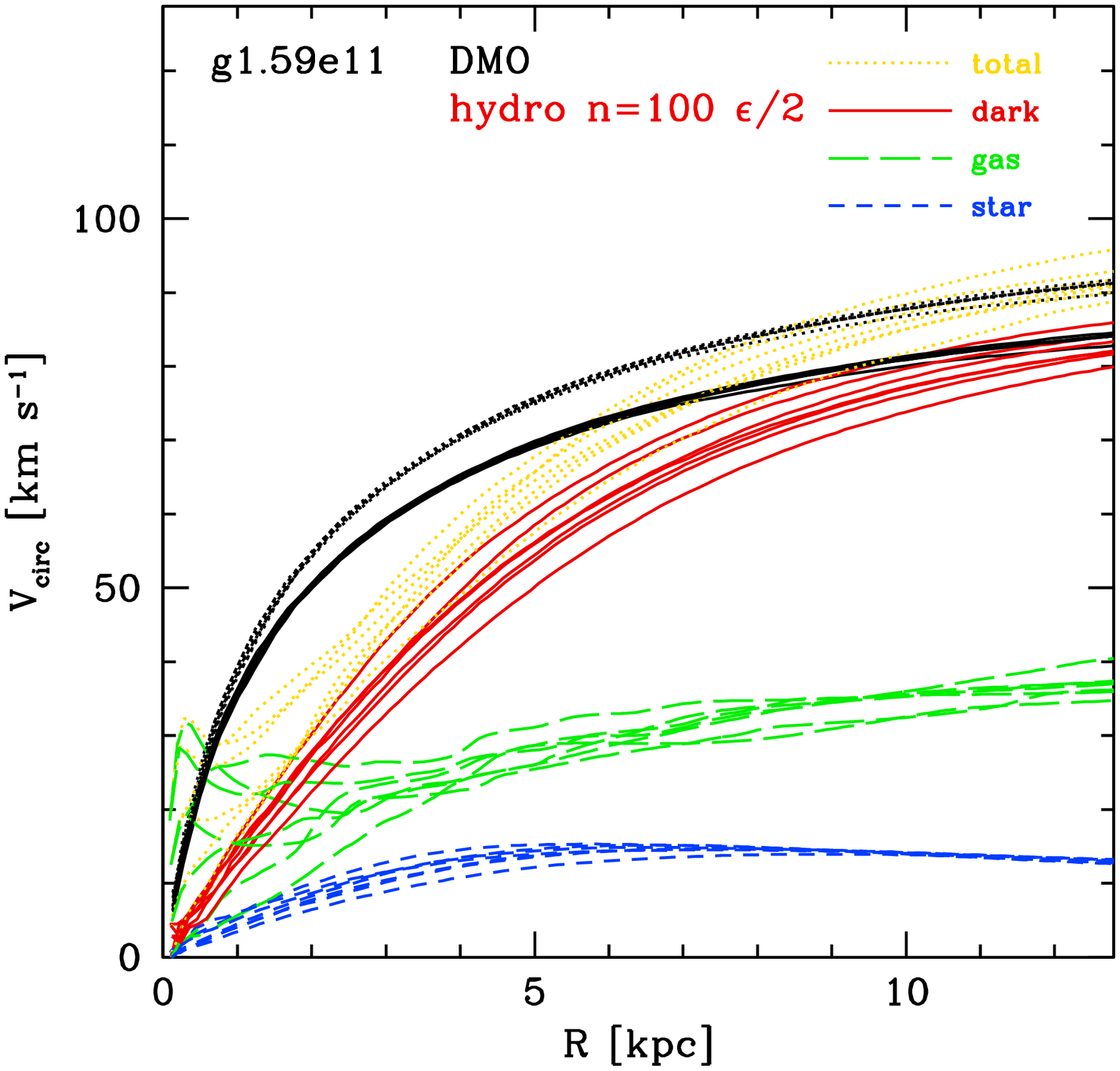}
  \includegraphics[width=0.45\textwidth]{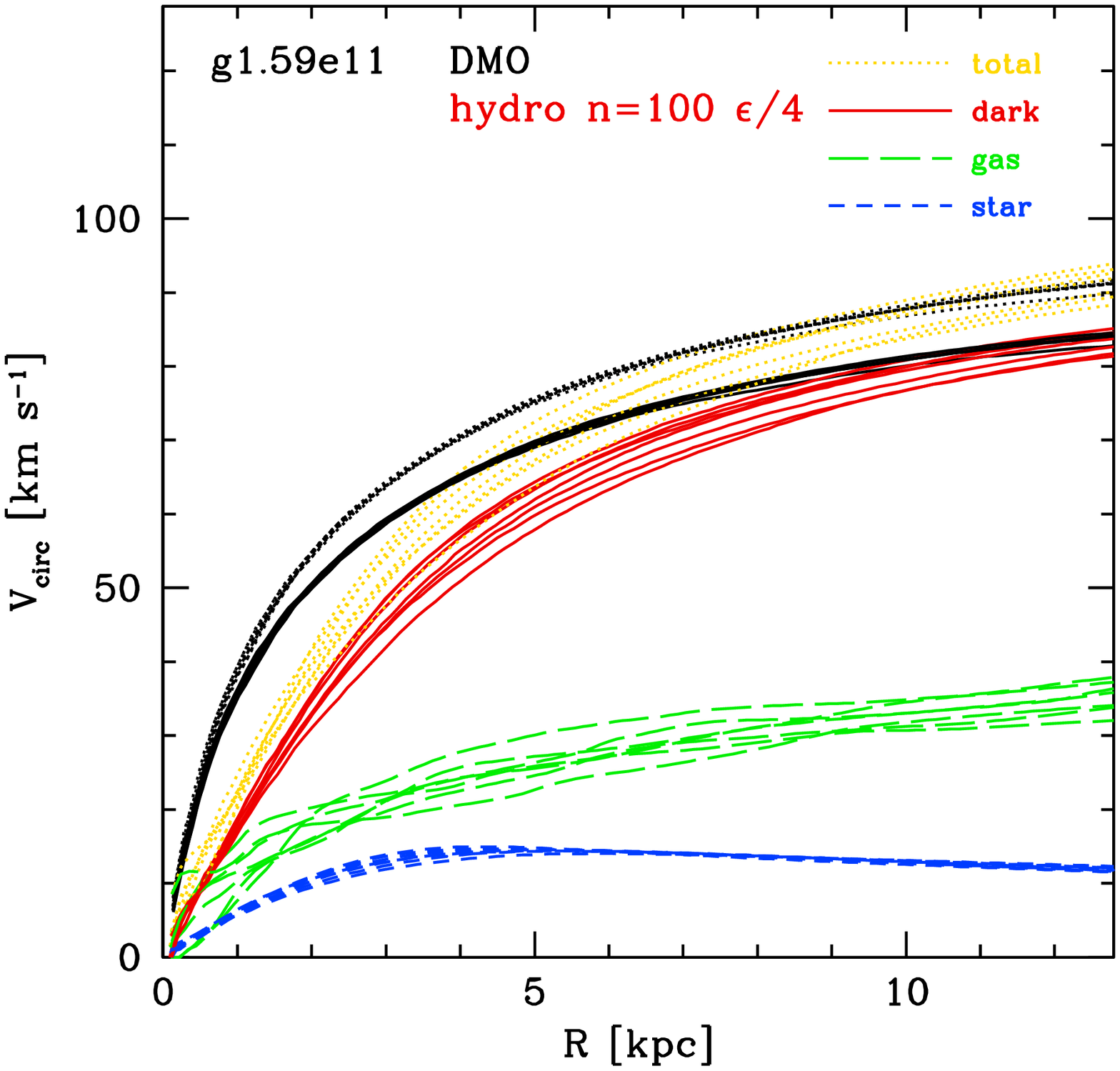}
  \includegraphics[width=0.45\textwidth]{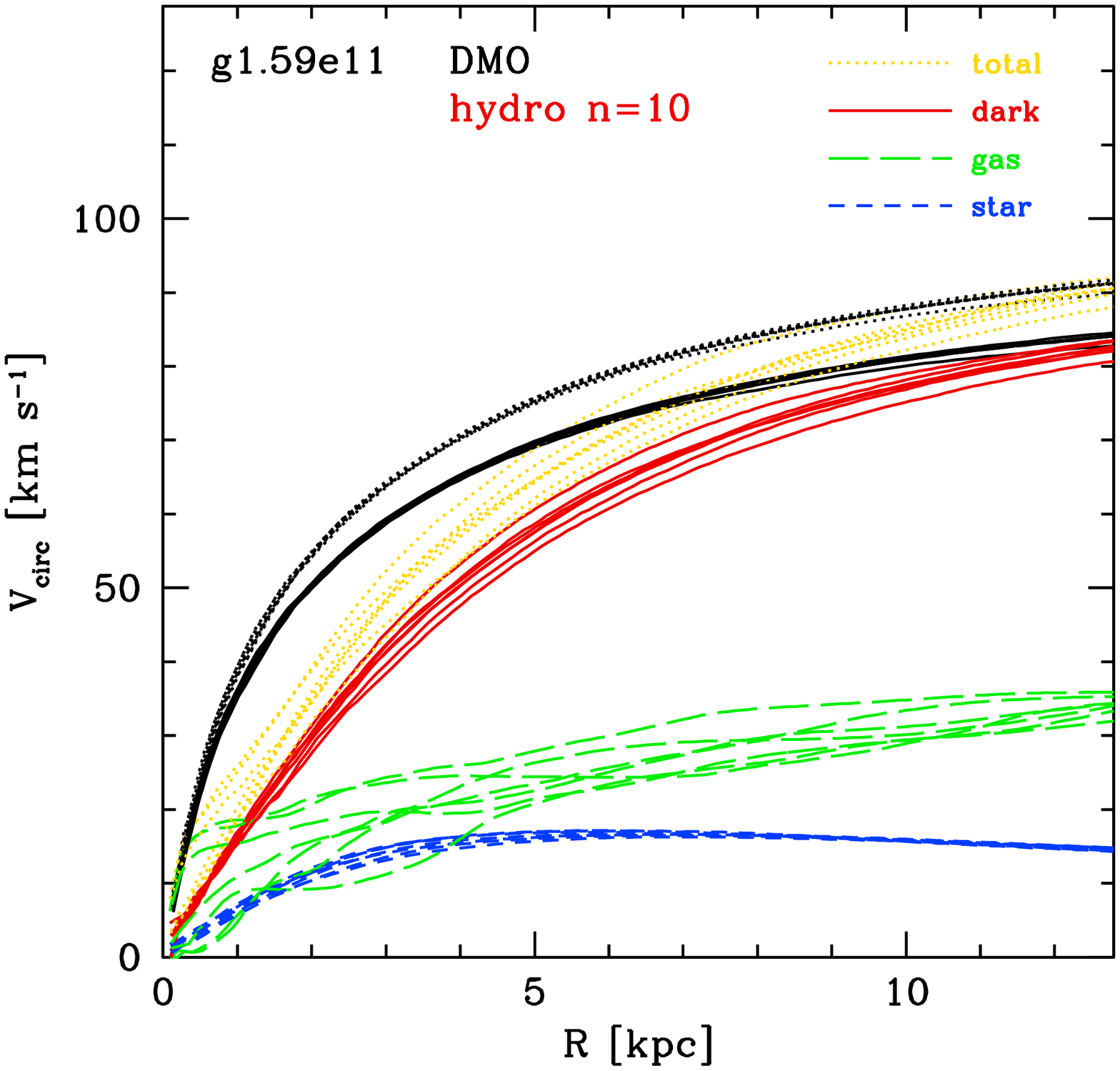}
  \caption{Effect of force softening on galaxy circular velocity
    profiles for g1.59e11. We show 7 outputs equally spaced in time
    between redshifts $z=0.5$ and $z=0.0$. In each panel the black
    lines show the DMO simulation: total (dotted) and scaled by
    $(\Omegam-\Omegab)/\Omegam$ (solid). The colored lines show the
    hydro simulation.  The DMO simulation has a very stable velocity
    profile. The $n=10$ (lower right), $n=100$ quarter softening
    (lower left), and $n=100$ half-softening (upper right) simulations
    have similar stellar (blue lines), gas (green lines), dark matter
    (red lines) and total (yellow lines) profiles. In particular, the
    dark matter profiles show only small variation and noticeable
    expansion with respect to the DMO. By contrast the $n=100$ (upper
    left) simulation has a high variability in the dark matter
    profiles at small radii, which can be traced to the high
    variability in the gas profile at small radii.}
  \label{fig:vr}
\end{figure*}

\section{Convergence in halo response at high star formation threshold}

The main result of this paper is shown in Fig.~\ref{fig:alpha}. This
shows the change in the dark matter mass profile at 1 per cent of the
virial radius (identical to the change in enclosed dark matter
density), while the lower panel shows the slope of the enclosed dark
matter density profile between 1  and 2 per cent of the virial
radius. Note that here we use the enclosed dark matter density rather
than the  local dark matter density as used in our previous works
\citep[e.g.,][]{Tollet16}, but the results are qualitatively the same
(compare the grey lines with the red circles).

In the upper panels, the dashed line corresponds to the DMO simulation
(by definition), while in the lower panels the open circles show the
DMO simulations (which cluster close to the NFW slope of $-1$).
Results are shown versus stellar-to-halo mass ratio as this parameter
has been shown to be better correlated with the halo response
\citep{DiCintio14, Dutton16b, Bullock17} than either the stellar mass
or halo mass alone.

For the lowest star formation threshold simulations $n=0.1$ (green
triangles) the haloes are essentially unchanged for
$\Mstar/M_{200}\lta 10^{-2}$, while they contract for higher
$\Mstar/M_{200}$ due to the increased dissipation of gas.  We show
three sets of simulations for $n=0.1$: the fiducial early stellar
feedback efficiency $\epsilon_{\rm ESF}=0.13$ (pronged green
triangles), $\epsilon_{\rm ESF}=0.04$ (solid green triangles), and
$c_{\ast}=0.015$ (open green triangles). All three sets of simulations
show the same trend of halo response with $\Mstar/M_{200}$.
These results for $n=0.1$ are very similar to those for the APOSTLE and
AURIGA simulations presented by \citet{Bose19}.
This agreement is encouraging given the numerous differences between
the codes. It suggests that how supernova feedback
is modeled as well as the hydrodynamical scheme are of secondary
importance compared to the star formation threshold.
Furthermore, we know of no cosmological galaxy formation
simulation that contradicts this result.  We thus conclude the lack of
halo expansion for simulations run with low star formation thresholds
$n\sim 0.1$ is a robust theoretical result. As discussed previously,
and also below, if low-mass CDM haloes do not expand then the
so-called too-big-to-fail problem for field galaxies represents a
serious challenge for the CDM model \citep{Garrison-Kimmel14,
  Dutton16a, Dutton19c}.

\begin{figure*}
  \includegraphics[width=0.45\textwidth]{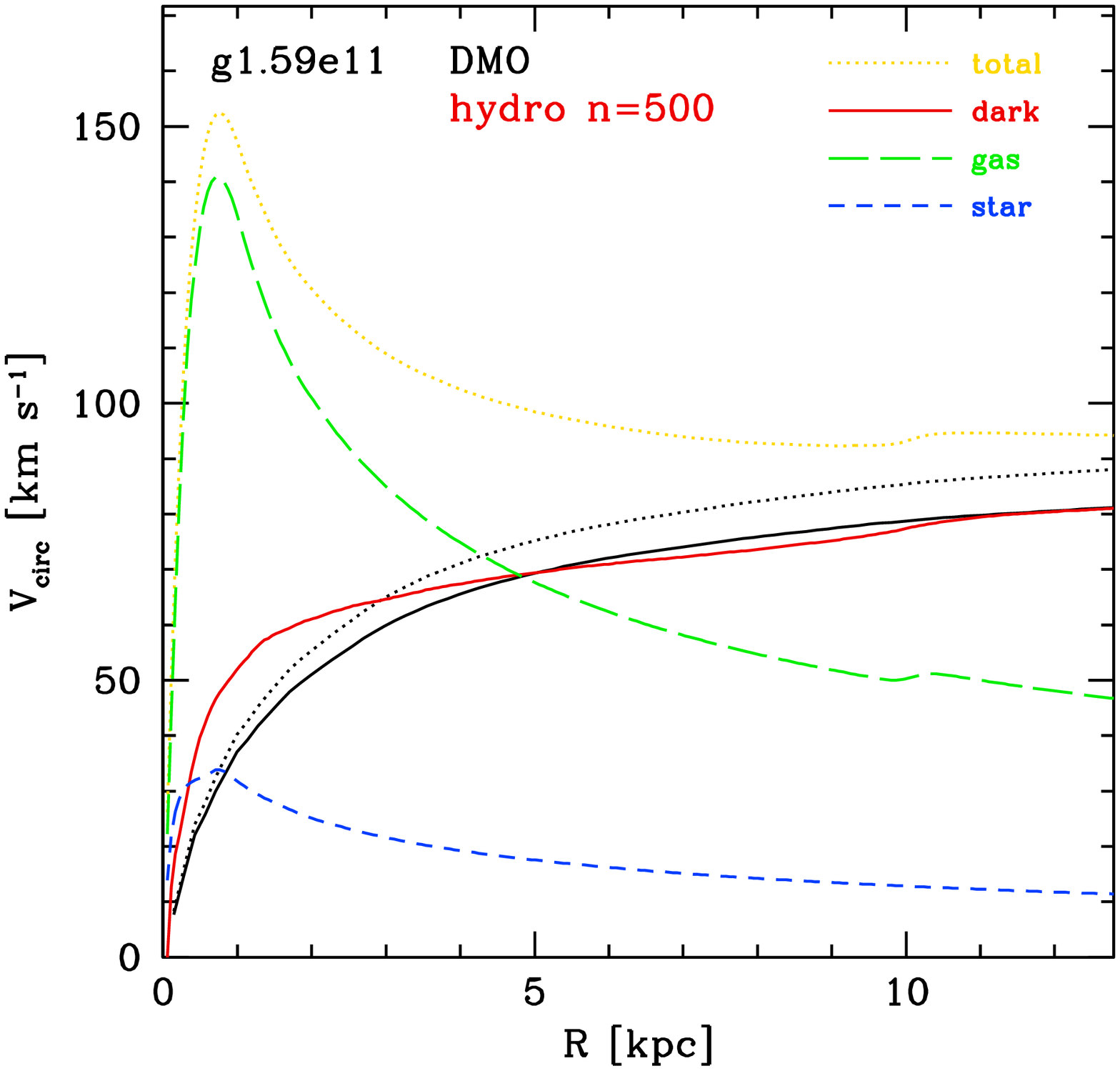}
  \includegraphics[width=0.45\textwidth]{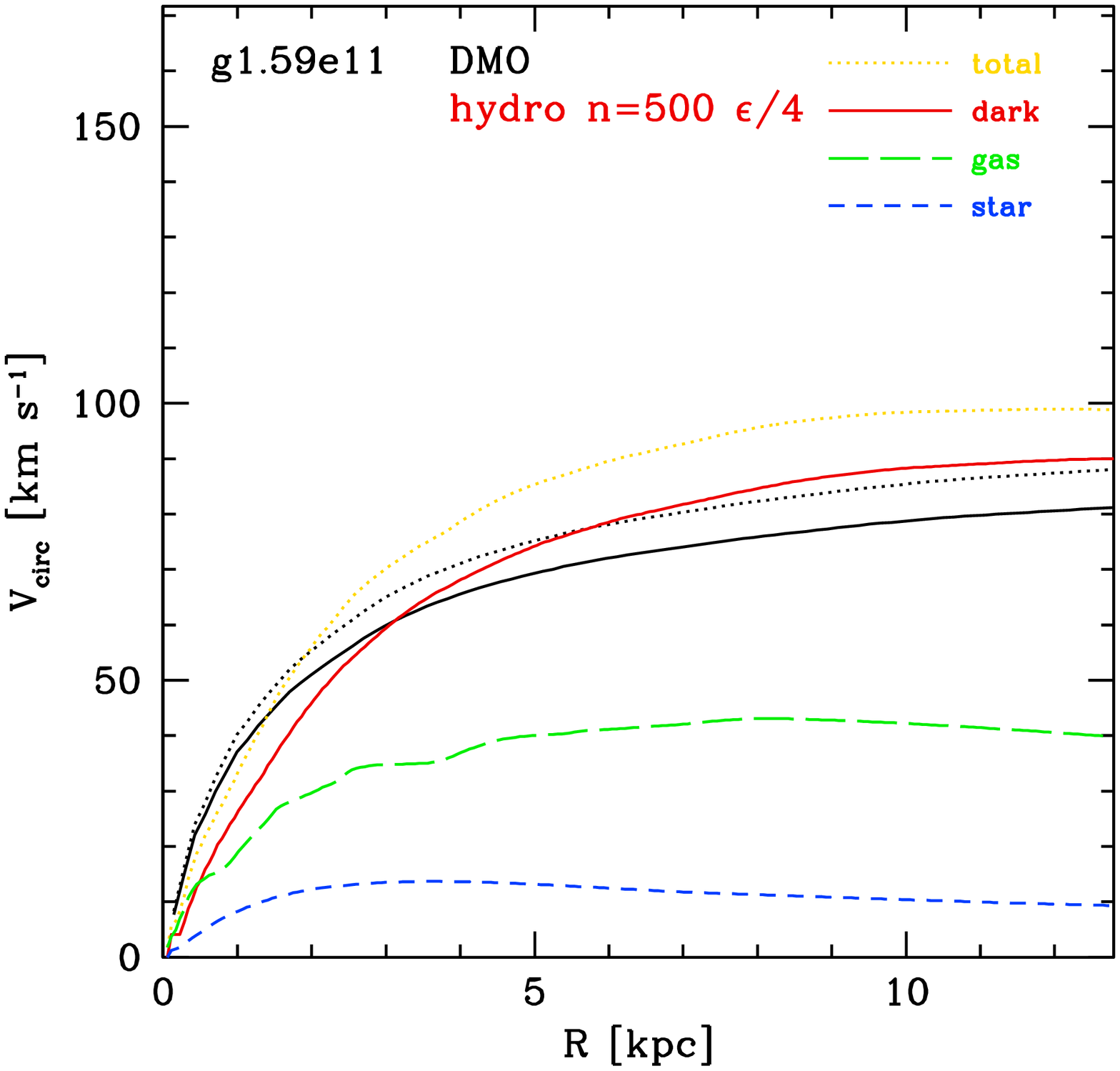}
  \caption{Effect of force softening on galaxy and halo mass profiles
    for halo g1.59e11 at $z=1.47$. Both simulations have
    a star formation threshold of $n=500$. With the fiducial force
    softening (left panel) the gas profile (green long-dashed) is very
    concentrated, and dominates the total mass within 4 kpc. This
    results in the dark matter profile (red solid) contracting with
    respect to the DMO simulation (black solid). With a force
    softening reduced by a factor of 4 (right panel), the gas profile
    is normal and the dark matter halo has started to expand within 3
    kpc.}
  \label{fig:vr2}
\end{figure*}

As we increase the star formation threshold to $n=1$ (blue squares)
the trends are similar to that from $n=0.1$, but with slightly
shallower DM slopes and slight halo expansion.  By $n=5$ (cyan
diamonds) strong halo expansion occurs when $10^{-3} \lta
\Mstar/M_{200} \lta 10^{-2}$, while haloes still contract for
$\Mstar/M_{200}\sim 10^{-1.5}$.
For high thresholds $n=10$ (red filled circles), $n=20$ (yellow
pentagons), $n=100$ (magenta hexagons), and $n=500$ (grey stars) the
halo response appears to have converged with strong expansion to dark
matter cores for $10^{-3} \lta \Mstar/M_{200} \lta 10^{-2}$.

The concept of convergence in halo response is complicated because in
our simulations halo expansion is primarily caused by feedback driven
gas outflows. These occur during bursts of star formation. The timing
of these bursts is not identical in different simulations run with
identical initial conditions due to the stochastic nature of star
formation in our simulations. Thus when we talk about convergence we
focus on quantities averaged over several time steps and/or
simulations with a given halo mass today.

The convergence in halo response at high star formation thresholds is
shown for four individual haloes in Fig.~\ref{fig:deltam}. The
vertical axis shows the ratio between enclosed dark matter masses in
the hydro and DMO simulations at 1 per cent of the virial radius
(i.e., the same parameter as in the upper panel of
Fig.~\ref{fig:alpha}).  For each simulation we show the results of
seven snapshots equally spaced in time between redshifts $z=0.5$ and
$z=0$. This shows that for a given halo there is significant
variability in the halo structure as a function of time.  This
variability is due to a combination of stochastic processes and
systematic evolution.  Nevertheless, if we look at the trends we see a
strong difference in the average halo response between $n=1$ (blue
squares) and $n=5$ (cyan diamonds). For $n\ge 10$ the halo response
shows good convergence for a given halo.  By convergence we mean the
average dark matter mass is independent of $n$.   The exception is
that for $n=100$ we see a larger variation in the halo response for
the fiducial softening simulations (open hexagons) versus the
half-softening simulations (filled hexagons). For g1.59e11 we even see
a few snapshots with mass ratios close to unity.  We trace the origin
of this feature in the $n=100$ simulations to insufficient spatial
resolution. When the star formation threshold is less than $n_{\rm
  max} \simeq 10 [{\rm cm}^{-3}]$ (for our fiducial resolution
simulations) gas can locally fragment and thus turn into stars, the
resulting feedback pushes gas out preventing a build up of gas in the
galaxy center. When the threshold is higher than $n_{\rm max}$,
instead of forming stars, the gas loses angular momentum and collapses
to the center of the galaxy until it is globally above the star
formation threshold. A large starburst and gas outflow event follows,
and then the process repeats.

An illustration of this is shown for halo g1.59e11 in
Fig.~\ref{fig:vr}.  Each panel shows the circular velocity profiles
for seven snapshots equally spaced in time between redshift $z=0.5$
and $z=0$. For our standard simulations $n_{\rm max}=10$ (lower right)
the stellar (blue), gas (green), and dark matter (red) profiles are
fairly stable, and the dark matter has expanded relative to the DMO
(black lines) show the total (dotted) and ``dark'' (solid) components.
For $n=100$ (upper left) there is a wide variation in the gas
profile. In some snapshots the gas is very concentrated, while in
others the gas has been blown out of the galaxy center. As would be
expected the dark matter variation follows the variation in the gas
(i.e., when the gas is more concentrated the dark matter is more
concentrated). In some snapshots the dark matter profile even follows
the DMO at small radii ($<$ kpc).  Since $n_{\rm max}\ll 100$ for our
fiducial choice of $m_{\rm gas}$ and $\epsilon_{\rm gas}$ we do not
expect this simulation to be physically realistic.

This problem with $n=100$ simulations can be simply fixed by reducing
the force softening of the particles.  When we reduce the force
softenings of the particles by a factor of 2 ($n_{\rm max}\simeq 80$,
upper right) and a factor of 4 ($n_{\rm max}\simeq 640$, lower left)
the resulting profiles (of the stars, gas, and dark matter) are very
similar to that of the $n=10$ simulations.  We note that these smaller
force softenings are within the bounds set by \citet{Ludlow19}, see
their fig.~1. Note also that there has been no recalibration of these
simulations.

Fig.~\ref{fig:vr2} shows the same halo but at an earlier time
($z=1.47$) and now with $n=500$.  The standard softening run (upper
left) now has an extremely overcooled gas bulge, which dominates the
central potential and has resulted in dark halo contraction (the solid
red line is above the solid black line below 5 kpc). The dense gas
bulge also makes the simulation much slower to run, which is why we
stopped it at high-redshift.  The $n=500$ quarter softening run
(right) has similar star, gas, and dark matter profiles to the $n=100$
quarter softening and $n=10$ simulations.  These simulations are all
still dark matter dominated at small radii, and the inner dark matter halo
has expanded compared to the DMO case.

\citet{Benitez19} showed that the inner dark matter content of
low-mass haloes (and the size of their cores) is very sensitive to the
assumed star formation threshold in the EAGLE model, hindering robust
model predictions and the interpretation of observational data.
The dwarf galaxy simulations of  \citet{Benitez19} have $m_{\rm
  gas}=6.6\times10^4~\Msun$ and $\epsilon_{\rm gas}=234$ pc, which
results in $n_{\rm max}=10.5$ (i.e., almost identical to our fiducial
simulations).  Their halo response is stable for $n=10$ to $n=80$. For
$n=160$ they start to see the effects of overcooling. By $n=320$ and
$n=640$ the gas dominates the central potential and the dark matter
halo contracts. Their results are thus completely consistent with what
we present here for our fiducial force softening runs.

\citet{Benitez19} suggest that it is the inefficiency of supernova
feedback at higher gas densities in the EAGLE code \citep{Crain15}
that is responsible for the increased central gas densities when
adopting a higher star formation threshold.  They suggest that the
numerical implementation of feedback will be important at high gas
densities since this (over-cooling) effect does not occur in the
FIRE-2 \citep{Hopkins18} galaxies which adopt a very high threshold of $n\sim 1000$.

Our results suggest a much simpler explanation, namely that the force
softening used by \citet{Benitez19} is not appropriate for star
formation thresholds significantly greater than $n\sim10$.  Indeed,
the FIRE-2 \citep{Fitts17} simulations are able to form galaxies with
normal looking gas profiles with $n\sim 1000$ simply because they
adopt very small force softenings ($\sim 20$ times smaller than
fiducial NIHAO for a given particle mass)\footnote{\citet{Fitts17} use
  $\epsilon_{\rm gas}=2$ pc for $m_{\rm gas}=500~\Msun$. Compare this to
  $\epsilon_{\rm gas}=89.4$ pc for $m_{\rm gas}=3474~\Msun$ for NIHAO,
  which scales to $\epsilon_{\rm gas}=46.9$ pc for $m_{\rm gas}=500~
  \Msun$.}.

\begin{figure*}
  \includegraphics[width=0.45\textwidth]{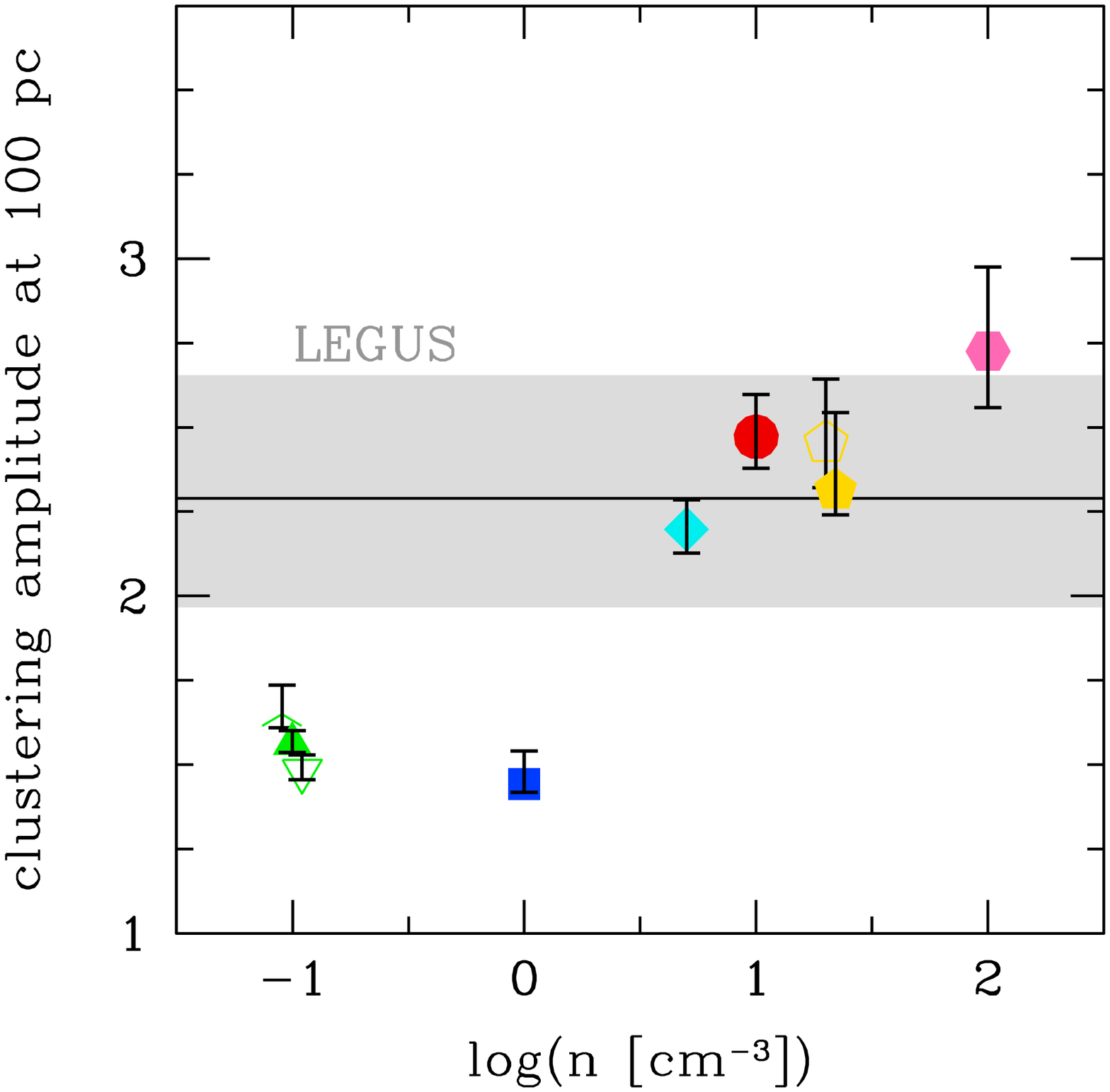}
  \includegraphics[width=0.45\textwidth]{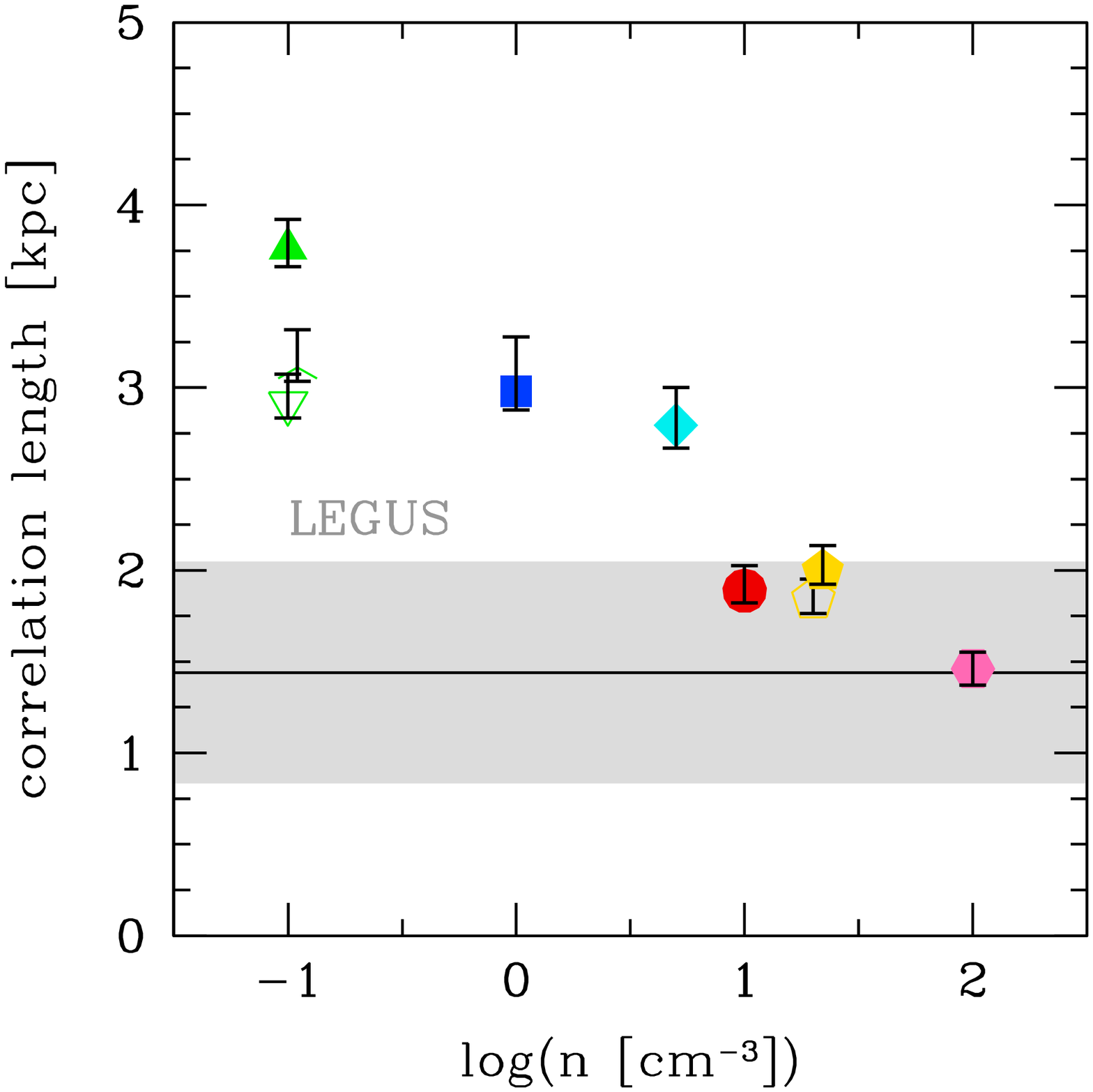}
  \caption{Clustering of young stars vs star formation threshold for
    galaxies with stellar masses $2\times 10^{8} \lta \Mstar \lta
    3\times 10^{10}~\Msun$. The left panel shows the clustering
    strength at 100 pc, $[ 1+\xi(r=100 {\rm pc})]$, while the right
    panel shows the    correlation length, $r(\xi=0)$. The shaded
    region shows observational results from LEGUS
    \citep{Grasha17}. Colored points show the simulation median with
    error bars indicating the uncertainty on the median. Low star
    formation thresholds ($n\le 1$) are disfavored by more than
    2$\sigma$.}
  \label{fig:tpcf}
\end{figure*}

To summarize, the response of dark matter haloes to galaxy formation
is sensitive to the star formation threshold. However, the dependence
on star formation threshold is quite simple, and is converged: at low
thresholds $n\lta 1$ dwarf galaxy haloes essentially follow the
DMO predictions, while for $n\gta 10$ haloes expand for $0.001 \lta
\Mstar/M_{200}\lta 0.01$.

\begin{figure*}
  \includegraphics[width=0.8\textwidth]{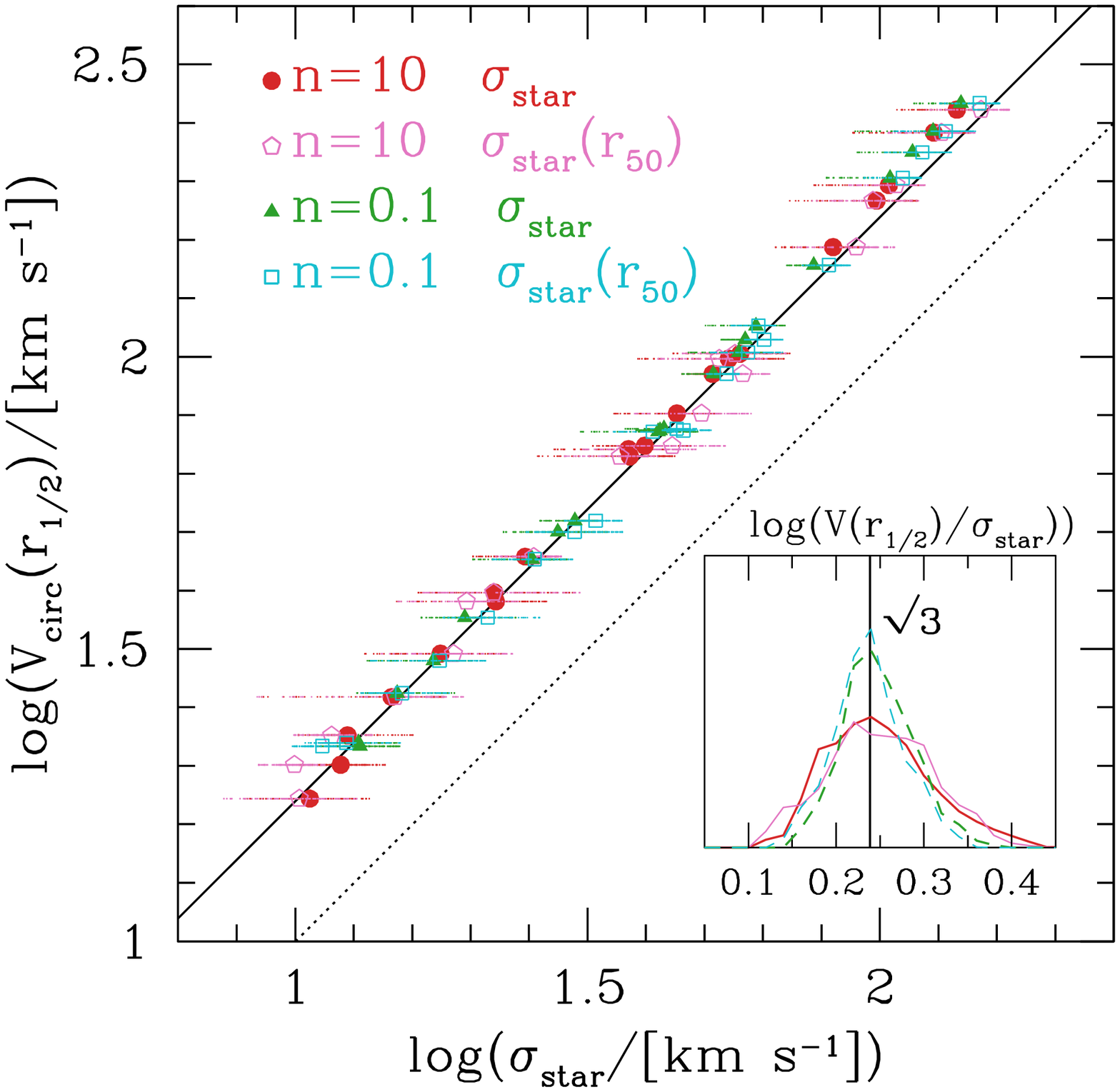}
  \caption{Circular velocity at the 3D half stellar mass radius vs the
    stellar velocity dispersion for simulations with $n=10$ and
    $n=0.1$. Small dots show single results from the 100 random
    projections.     Large symbols show the median value for each
    galaxy. Filled symbols use the velocity dispersion of all the
    stars (within $0.2 R_{200}$), open symbols use the velocity
    dispersion of the stars within the projected circular half stellar
    mass radius, $R_{50}$. The dotted line shows the 1:1 relation. The
    solid line shows the prediction from the virial theorem of $V_{\rm
      circ}(r_{1/2})=\sqrt{3}\sigma_{\rm star}$, which is a good
    approximation in our simulations.  The inset panel shows
    histograms of the ratio $V(r1/2)/\sigma$ which is approximately
    log-normally distributed with a standard deviation of about 0.04
    (for $n=0.1$) to 0.06 ($n=10$) dex.}
  \label{fig:vsigma}
\end{figure*}

\section{Constraining the star formation threshold with observations}

Having established how the structure of dark matter haloes depends
strongly on the star formation threshold we now turn to observations
to calibrate this free parameter. Since our ultimate goal is to use
these calibrated simulations to test the CDM model, we want to use
observations that are not directly related to the structure of dark
matter haloes.

\citet{Buck19b} showed that the clustering of young stars in the NIHAO
simulations depends strongly on the adopted star formation
threshold. The clustering can be quantified using the
two-point-correlation statistic of young star particles.  Here we
repeat the analysis of \citet{Buck19b} using more values of the star
formation threshold ($n=5$, $n=100$). To match the stellar mass range
of the observed galaxies in \citet{Grasha17} we take the 8 simulations
with redshift $z=0$ stellar masses $2\times 10^8 \lta \Mstar \lta
3\times 10^{10}~\Msun$. We use 24 snapshots evenly spaced in time from
$z=0.5$ to $z=0.0$.  For each snapshot we calculate the
two-point-correlation statistic vs separation and fit the data with a
function \citep[see][for details]{Buck19b}.
From the fit we calculate the clustering amplitude at a
separation of 100 pc, and the radius where the clustering amplitude is
equal to unity (i.e., that of a random distribution). For each set of
simulations with a given $n$ we find the median and scatter of the
$\sim 200$ outputs.

Fig.~\ref{fig:tpcf} shows the clustering amplitude at 100 pc (left)
and the correlation length (right) versus the star formation threshold
of the simulation.  The grey bands show the 1$\sigma$ region of the
observations using data from the LEGUS survey presented in
\citet{Grasha17}. Note that in making this comparison we are assuming
a correspondence between the clustering of observed young star
clusters and the clustering of young simulated star particles. For the
observations we use results for ages less than 40 Myr and all classes
1,2,3. In the simulations we only apply an age cut.  The star
particles in our simulations have similar masses as the observed star
clusters ($\sim 10^3$ to $\sim10^4 \Msun$). Thus to a first order
approximation our assumption that young star particles trace young
stellar clusters seems reasonable. However, an in depth investigation
into the correspondence between the clustering of simulated star
particles and observed star clusters, and ways to reduce any biases is
certainly warranted.

For each simulation the symbol shows the median value, while the
error-bar shows the error on the median ($1/\sqrt{N}$ times the
standard deviation).  As expected there is a clear trend for stronger
clustering (higher clustering amplitude and smaller correlation
length) with higher star formation thresholds.  The simulations with
$n=5,10,20,100$ overlap with the observed clustering, while the
simulations with $n=0.1$ and $n=1$ are more than $2\sigma$ away from
the observations, with clustering that is too weak.  The multiple
points for $n=0.1$ and $n=20$ show that the clustering strength is not
sensitive to the feedback or star formation efficiency. 

In summary, the clustering of young stars provides strong constraints
to the sub-grid model for star formation in our simulations.  Star
formation thresholds of $n\le 1$ are strongly disfavored, while $10
\lta n \lta 100$ provides a good match to the observed clustering.
Even though the clustering does not single out a specific value for
the star formation threshold, it still provides a very useful
constraint because a majority of the galaxy formation simulations in
the literature adopt a low star formation threshold ($n\sim
0.1$). Furthermore, as shown in \citet{Dutton19c} and below, the inner
structure of the dark matter halo also depends on the star formation
threshold.

\begin{figure*}
  \includegraphics[width=0.9\textwidth]{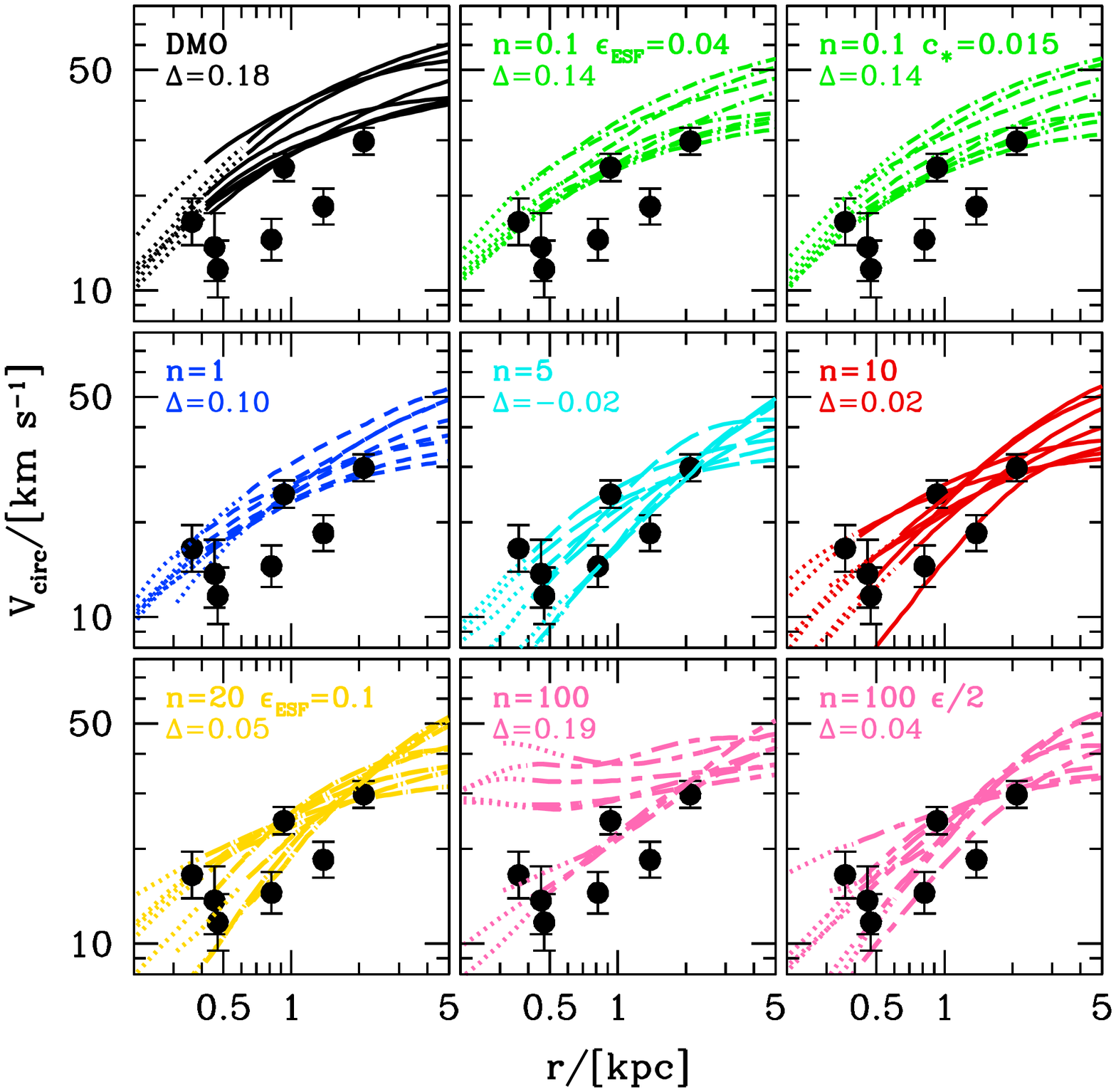}  
  \caption{Circular velocity versus radius for dwarf galaxies with
    stellar masses $10^6 \lta \Mstar \lta 10^8~\Msun$.   Filled circles
    with error bars show observed field galaxies more than 500 kpc
    from the Milky Way \citep{Kirby14}.   Lines show our simulations,
    where the  transition to dotted lines marks the scale that is
    accurately resolved (twice the dark matter softening).  The
    parameter $\Delta$ is the mean offset (in dex) between the
    observations and the simulations. The DMO simulations (upper left)
    are offset from the observations by an average of 0.18 dex (i.e.,
    a factor of 1.5). As the star formation threshold increases the
    offset decreases, such that with $n=10$ (middle right) the offset
    is just 0.02 dex.}
  \label{fig:tbtf}
\end{figure*}

\section{Testing CDM with galaxy kinematics}

Having established how the structure of CDM haloes depends on the star
formation threshold, and calibrated this free parameter using the
clustering of young stars, we now turn to observations of galaxy
circular velocities to test the resulting predictions for CDM.
Following the analysis in \citet{Dutton19c} we split the tests into
two mass ranges corresponding to dwarf galaxies ($10^6 \lta
\Mstar/~\Msun \lta 10^8$) and intermediate-mass galaxies ($10^9 \lta
\Mstar/~\Msun \lta 10^{10}$). This split is based on the availability
of dynamical tracers. For the low-mass galaxies we use the integrated
stellar velocity dispersions to trace the circular velocity within the
half-light radius, while for the intermediate-mass galaxies we use
resolved rotation curves.

First, we show that the projected stellar velocity dispersion,
$\sigma_{\rm star}$, is a good tracer of the circular velocity at the
3D stellar half-mass radius, $r_{1/2}$. Fig.~\ref{fig:vsigma} shows
the relation between circular velocity and velocity dispersion for all
20 simulated galaxies at $z=0$ using 100 random projections per
galaxy. Red circles and magenta pentagons show results for $n=10$,
while green triangles and cyan squares show results for $n=0.1$ (and
re-calibrated with $\epsilon_{\rm ESF}=0.04$).  Red and green points
show the stellar velocity dispersion measured within the whole galaxy
(defined to be 0.2 virial radii, $R_{200}$) while the magenta and cyan
points show the stellar velocity dispersion within the projected
stellar half-mass radius.  Fig.~\ref{fig:vsigma} shows that the ratio
between circular velocity and stellar velocity dispersion is, on
average, insensitive to the star formation threshold of the
simulation, and the aperture within which the velocity dispersion is
measured.

On average we find $V_{\rm circ}(r_{1/2})=\sqrt{3}\sigma$, as predicted by the
spherical Jeans equations \citep{Wolf10}. There is a
non-negligible scatter of $\simeq 0.06$ dex in this relation.  The
galaxy to galaxy variation is relatively small. Most of the scatter
comes from variations resulting from different viewing angles.  Thus
samples of galaxies are needed to bring down the sampling errors.

\subsection{Dwarf galaxies}

For the 8 lowest mass haloes in our sample in Fig.~\ref{fig:tbtf} we
compare the simulated circular velocity profiles (lines) to the
circular velocity at the 3D half-light radius of observed field dwarf
galaxies in the Local Group (points with error bars) from
\citet{Kirby14}. The simulated galaxies have stellar masses in the
range $10^6 \lta \Mstar \lta 10^8 ~\Msun$, while the observed dwarfs
have with V-band luminosities from $10^6$ to $2\times 10^8 \Lsun$. In
addition, for the observations we have excluded galaxies with
distances less than 500 kpc ($\sim 2$ virial radii) from the Milky Way
to minimize contamination of back-splash galaxies \citep{Buck19a}.
Note that four panels in Fig.~\ref{fig:tbtf} have previously been
published in fig.~4 of \citep{Dutton19c}. The reproduced panels are:
DMO (top left), $n=0.1$ $\epsilon_{\rm ESF}=0.04$ (top center), $n=1$
(middle left), and $n=10$ (middle right). Here we reproduce these
results and include five additional sets of simulations: $n=0.1$
$c_{\ast}=0.015$ (top right), $n=5$ (middle center), $n=20$ (lower
left), $n=100$ (lower center), and $n=100$ with half force-softening
(lower right).

As in \citet{Dutton19c}, we calculate the average offset between the
observations ($V_{\rm obs}$) and simulations ($V_{\rm sim}$).  For
each observed data point, $V_{{\rm obs},i}(r_{{\rm obs},i})$, the mean
offset at radius $r_{{\rm obs},i}$ with respect to the $N_{\rm sim}=8$
simulations is
\begin{equation}
\label{eq:delta}
  \Delta_i=\sum_{j=1}^{N_{\rm sim}}(\log_{10} V_{{\rm obs},i}(r_{{\rm obs},i}) -\log_{10} V_{{\rm sim},j}({r_{\rm obs},i})/N_{\rm sim}.
\end{equation}
We then take the mean of $\Delta_i$ over the 7 observed data points,
which we denote $\Delta$.

\begin{figure*}
  \includegraphics[width=0.9\textwidth]{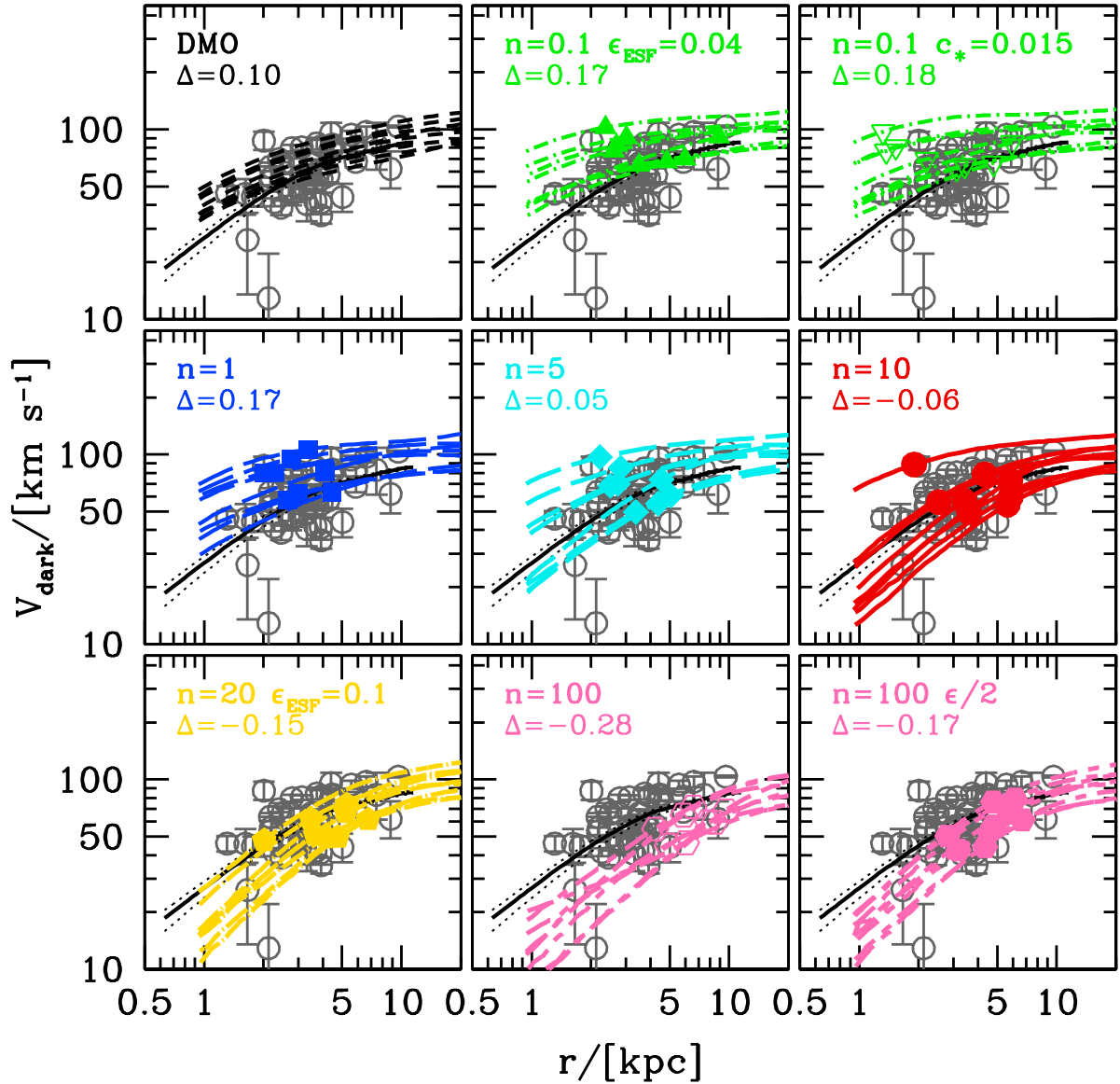}
  \caption{Dark matter circular velocity versus radius for galaxies
    with $10^{8.8} \lta \Mstar \lta 10^{10.2}~\Msun$.  The same observations
    are shown in all panels. Grey circles with error-bars show the
    dark matter circular velocity at the half-light radius for
    observed galaxies from SPARC.  The    solid black line shows the
    mean dark matter circular velocity curve of the observations,
    while the dotted lines show the 1 $\sigma$ scatter.
    Each panel shows a different set of
    simulations with star formation threshold $n$ as indicated.
    The points are located at the projected stellar half-mass radii .  The
    value $\Delta$ is the mean offset [dex] between the simulations
    and the observed average velocity at 2 kpc.}
  \label{fig:vr_sparc}
\end{figure*}

As previously shown in fig.~4 from \citet{Dutton19c} we see that the
DMO simulations (upper left panel) are systematically too high.  The
mean offset $\Delta=0.18$, i.e. the average offset between simulation
and observation is a factor of 1.5 in velocity, and a factor of 2.3 in
enclosed mass.  This recovers the well known too-big-to-fail problem
of Local Group field galaxies \citep{Garrison-Kimmel14}.

The lower threshold
hydro simulations $n=0.1$ ($\Delta=0.14$) and $n=1$ ($\Delta=0.10$)
can reproduce some, but not all of the observed data points, and
predict circular velocities that are systematically too high.  In
particular for $n=0.1$ the two sets of simulations show that the dark
matter profiles are insensitive to the efficiencies of early stellar
feedback and star formation.

We showed previously in \citet{Dutton16a,Dutton19c} that the NIHAO
simulations resolve the too-big-to-fail problem. The fiducial NIHAO $n=10$
simulations match the observations well with $\Delta=0.02$ (middle
right panel). The new simulations ($n=5$, recalibrated $n=20$, and
half-softening $n=100$) also provide a good match to the observations
which is interesting as these are the star formation thresholds that
are consistent with the observed clustering of young stars shown above
in Fig.~\ref{fig:tpcf}.
Notice that the $n=100$ standard softening simulations result in a
high $\Delta=0.19$.  This is due to the build up of gas in the galaxy
centers (because the gas cannot easily reach the star formation
threshold), rather than a strong contraction of the dark matter halo,
or a significant stellar component. See the upper left panel in
Fig.~\ref{fig:vr} for an example.

\subsection{Intermediate-mass galaxies}

We now consider the simulated galaxies with stellar masses in the
range $10^9 \lta \Mstar \lta 10^{10} ~\Msun$. As with the dwarf
galaxies there are 8 haloes in this mass range. In
Fig.~\ref{fig:vr_sparc} we compare the simulated circular velocity
profiles (lines) to the dark matter circular velocity at the
half-light radius (grey circles with error bars) from the SPARC survey
of nearby star forming galaxies \citep{Lelli16}.  This figure extends
the results previously shown in fig.~5 from
\citet{Dutton19c}. Specifically the following panels are reproduced
from \citet{Dutton19c}: DMO (top left), $n=0.1$ $\epsilon_{\rm
  ESF}=0.04$ (top center), $n=1$ (middle left), and $n=10$ (middle
right).

For observations the dark matter circular velocity is  obtained by
subtracting the stellar and gas circular velocity profile from the
total rotation velocity, assuming a stellar mass-to-light ratio at
$3.6\,\mu$m of 0.5.  The solid black line shows the average velocity
profile of the observations plotted between the average smallest and
largest point on the rotation curve.  Because these galaxies tend to
be dark matter dominated, there is only a small uncertainty in the
dark matter profile caused by the $\sim 0.1$ dex uncertainty in
stellar mass-to-light ratio (dotted lines).  Larger uncertainties are
how accurately the rotation curve corrected for inclination traces the
circular velocity, and sampling effects since SPARC is not a volume
limited survey.

The colored lines show the simulated dark matter circular velocity
profiles computed in spheres: $V_{\rm circ}=\sqrt{(G M(<r)/r)}$. For
the DMO simulations the total profile has been rescaled by the cosmic
baryon fraction ($\sqrt{1-\fbar}\simeq 0.92$).  For the hydro
simulations symbols are located at the projected half-mass radius of
the stars. This shows that the galaxy sizes for these simulations are
in reasonable agreement with the observations and that there is only a
small dependence of the sizes on the star formation threshold.

The parameter $\Delta$ (see Eq.~\ref{eq:delta}) is computed at a
radius of 2 kpc.  This is chosen as it is the smallest spatial scale
that is reliably resolved in both the simulations and observations for
all galaxies.  As with the dwarf galaxies in Fig.~\ref{fig:tbtf} we
see that the DMO, $n=0.1$ and $n=1$ simulations are systematically too
high, and as $n$ increases the dark matter velocities are
systematically reduced.  The simulations that best match the observed
clustering of young stars ($n\sim 10$), also provide the best match to
the observations of dark matter circular velocities.  Note that for
$n=100$ the under predicting of the observations is at least partly
due to these simulations under producing stars (see
Fig.~\ref{fig:msmv}). The $n=100$ $\epsilon/2$ simulations do match
the stellar masses for the dwarf galaxies, and also the velocities. So
the apparent failure of the $n=100$ $\epsilon/2$ simulations to match
the velocities in Fig.~\ref{fig:vr_sparc} should not be considered a
fatal failure of $n=100$ simulations.  A recalibration  would thus be
desirable before stronger conclusions for $n=100$ are drawn.

\begin{figure}
  \includegraphics[width=0.45\textwidth]{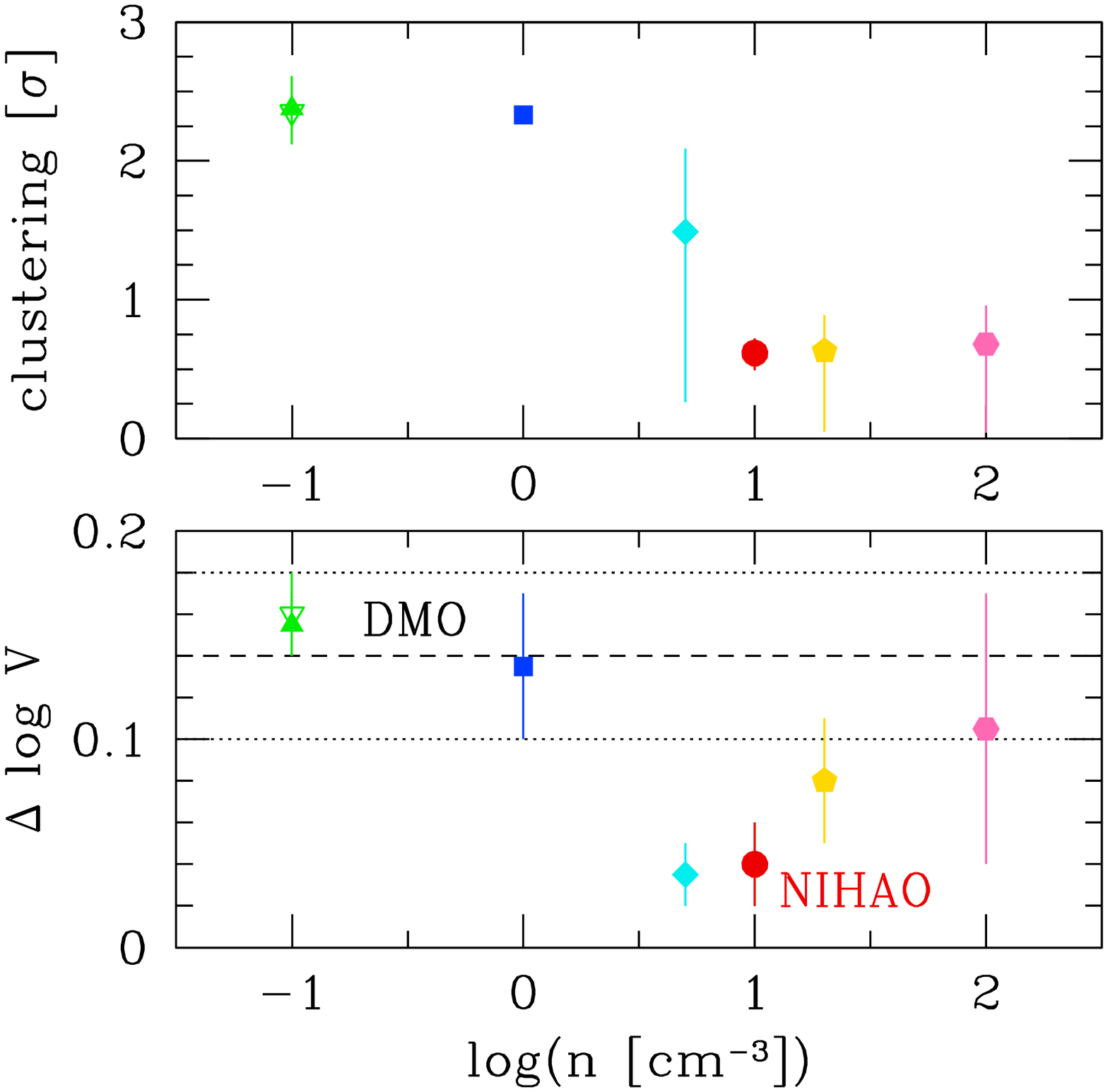}
  \caption{Summary of our results on the clustering of young stars
    (upper panel) and the circular velocities of nearby galaxies
    (lower panel) as a function of the star formation threshold of the
    simulations. The vertical error-bars show the range for two different
    measurements. In the lower panel the horizontal lines show the results for DMO simulations.}
  \label{fig:sum}
\end{figure}

A summary of the clustering and kinematic results is shown in
Fig.~\ref{fig:sum}. All the simulations shown here (except the $n=100$
at intermediate masses) match the stellar mass vs halo mass relation
(Fig.~\ref{fig:msmv}). The clustering measurements are used to provide
further calibration.  The upper panel shows the mean offset between
the simulated and observed clustering (from Fig.~\ref{fig:tpcf}) in
units of the observed uncertainty. The error-bar connects the two
measurements of clustering: amplitude at 100pc and correlation
length. Except for $n=5$ the two measurements give very similar
results. The $n=0.1$ and $n=1$ simulations do not reproduce the
observed clustering of young stars, and thus should not be used to
test CDM. Simulations with $n=10$ to $n=100$ reproduce the clustering
equally well.  The $n=5$ is consistent with only one measurement of
clustering, making this threshold borderline successful.

The lower panel shows the mean offset in $\log V$ between simulations
and observations ($\Delta$ parameter from Figs~\ref{fig:tbtf} \&
\ref{fig:vr_sparc}). This is used to test the CDM model. Again the
vertical error-bars connect the two measurements for dwarf and
intermediate-mass galaxies. The horizontal lines show the velocity
offsets for the DMO simulations, which are quite close to the values
for hydro simulations with $n=1$ and $n=0.1$.  Thus DMO and hydro
simulations with $n=0.1$ and $n=1$ fail to match observations. This is
not a problem for CDM because these simulations have already been
disfavored by the clustering measurements. 

The red circle shows the fiducial NIHAO $n=10$ which provides the best
match to the clustering of young stars and circular velocities of
nearby galaxies. We thus have shown that simulations that are
calibrated to reproduce both the stellar mass vs halo mass relation,
and clustering of young stars, have dark matter on small scales
consistent with CDM.


\section{Summary}

In this paper we investigated the impact of the star formation
threshold, $n$, on the response of the dark matter halo to galaxy
formation. Extending the study of \citet{Dutton19c} that looked at
$n=0.1,1,10$ here we consider thresholds as high as $n=500$. As with
\citet{Dutton19c} we use 20 sets of cosmological hydrodynamical
simulations from the NIHAO project \citep{Wang15} that simulate dark
matter haloes in the range $10^{10} \lta M_{200} \lta 10^{12} ~\Msun$
at redshift $z=0$.  We summarize our results as follows:

\begin{itemize}
  
\item We confirm the results of previous studies that the response of the
dark matter halo to galaxy formation is primarily a function of two
parameters: 1) the ratio between stellar mass and halo mass
\citep{DiCintio14, Chan15, Tollet16}, and 2) the adopted star formation
threshold, $n$, of the simulation \citep{Dutton19c, Benitez19}.

\item For high star formation thresholds ($n=5$ to $n=500$) the halo
  response has converged (Fig.~\ref{fig:alpha}), provided the
  simulation has sufficient spatial resolution to resolve the
  fragmentation of gas to densities above the gas density threshold in
  question.

\item We trace previous claims by \citet{Benitez19} for halo
  contraction at $n\gta 200$ to insufficient spatial resolution. With
  our default force softening the maximum gas density we can resolve
  is $n_{\rm max}\sim 10$. Applying our formula to the
  \citet{Benitez19} simulations also yields $n_{\rm max}=10.5$.  For
  our $n=100$ and $n=500$ simulations the gas is unable to locally
  fragment, instead it loses angular momentum and collapses to the
  center of the galaxy. The gas dominates the central potential
  causing the dark matter halo to contract (Fig.\ref{fig:vr2}).
  However, if we reduce the force softening of the gas by a factor of
   2 to 4, we recover the halo expansion achieved from our fiducial
  $n=10$ simulations using $n=100$ (Fig.~\ref{fig:vr}).

\item Following \citet{Buck19b} we use the spatial clustering of young
  stars to calibrate the star formation threshold parameter. The
  clustering strength increases roughly monotonically with star
  formation threshold (Fig.~\ref{fig:tpcf}).  Low threshold
  simulations ($n=0.1$ to $n=1$) are inconsistent with observations at
  more than $2\sigma$, while simulations with $n=10$ to $n=100$ are
  consistent with observations (Fig.~\ref{fig:sum}).
    
\item Finally, to test the CDM model we use the circular velocity vs
  radius plot for galaxies with stellar masses $10^6 \lta \Mstar \lta
  10^{10}$ (Figs.~\ref{fig:tbtf} \& \ref{fig:vr_sparc}).  DMO
  simulations and low star formation threshold simulations ($n=0.1, n=1$) fail by
  predicting a factor of $\simeq 2$ more mass than is
  observed. Simulations with $n\sim 10$ provide a good match to the
  observations.  

\item Investigating systematic effects in this test (for low-mass
  galaxies) we show that stellar velocity dispersions are an unbiased
  tracer of the circular velocity at the 3D stellar half-mass radius
  (Fig.\ref{fig:vsigma}).

\end{itemize}

We thus conclude that the CDM model provides a good description of the
structure of galaxies on kpc scales (once the effects of baryons are
properly taken into account).

With our choice of sub-grid models, only simulations with a high star
formation threshold can make accurate predictions for the structure of
CDM haloes. As well as the fiducial NIHAO \citet{Wang15} simulations
several other groups also adopt a high star formation threshold: e.g.,
\citet{Governato10,Governato12}, FIRE \citet{Hopkins14, Hopkins18},
and \citet{Read16}.  If this result extends to other choices of
sub-grid models, it means that a large fraction of the zoom-in
simulations in the literature: e.g., APOSTLE \citep{Sawala16}, AURIGA
\citep{Grand17}; and all of the large volume simulations: e.g., EAGLE
\citep{Schaye15}, ILLUSTRIS \citep{Vogelsberger14, Pillepich19},
ROMULUS \citep{Tremmel17} need to revise and re-calibrate their models
for star formation and feedback if they are to make accurate
predictions for the structure of cold dark matter haloes on kpc
scales. An additional consideration is that simulations with higher
star formation thresholds take significantly longer to run (due to the
higher gas densities, and subsequent smaller required time steps). Many
of the large volume simulations would not be computationally feasible
at present if run with a high star formation threshold model. Thus
there is a trade-off between the accuracy of the simulation and the
number of haloes that can be simulated.

Looking to the future, both the calibration of the star formation
threshold and the testing of CDM that we present are based on small
samples of simulated and observed galaxies. Thus significant
improvements in the accuracy are possible with larger samples.
Improved accuracy will enable the framework we have presented to be
applied to other models for dark matter, such as warm dark matter and
self-interacting dark matter.

  \section*{Acknowledgments}
We thank the anonymous referee for providing a constructive report
that improved the clarity of the paper.  This research was carried out
on the High Performance Computing resources at New York University Abu
Dhabi; on the  {\sc theo} cluster of the Max-Planck-Institut f\"ur
Astronomie and on the {\sc hydra} clusters at the Rechenzentrum in
Garching.
The authors gratefully acknowledge the Gauss Centre for Supercomputing
e.V. (www.gauss-centre.eu) for funding this project by providing
computing time on the GCS Supercomputer SuperMUC at Leibniz
Supercomputing Centre (www.lrz.de).
TB acknowledges support from the European Research Council under
ERC-CoG grant CRAGSMAN-646955.  AO is funded by the Deutsche
Forschungsgemeinschaft (DFG, German   Research Foundation) -- MO
2979/1-1.

 \section*{Data Availability Statement}
The data underlying this article will be shared on reasonable request
to the corresponding author.


\end{document}